\documentclass[fleqn,usenatbib]{mnras}

\usepackage{newtxtext,newtxmath}

\usepackage[T1]{fontenc}

\DeclareRobustCommand{\VAN}[3]{#2}
\let\VANthebibliography\thebibliography
\def\thebibliography{\DeclareRobustCommand{\VAN}[3]{##3}\VANthebibliography}

\usepackage{graphicx}	%
\usepackage{amsmath}	%
\usepackage{MnSymbol}	%

\title[Local parameterization of IGM damping wings]{A local, topology-independent parameterization of quasar IGM damping wings}

\author[Kist et al.]{
Timo Kist,$^{1}$\thanks{E-mail: kist@strw.leidenuniv.nl}
Joseph F. Hennawi$^{2,1}$
and Frederick B. Davies$^{3}$
\\
$^{1}$Leiden Observatory, Leiden University, P.O. Box 9513, 2300 RA Leiden,
The Netherlands\\
$^{2}$Department of Physics, University of California, Santa Barbara, CA 93106, USA\\
$^{3}$Max-Planck-Institut für Astronomie, Königstuhl 17, 69117 Heidelberg, Germany
}

\date{Accepted XXX. Received YYY; in original form ZZZ}

\pubyear{2024}

\begin{document}
\label{firstpage}
\pagerange{\pageref{firstpage}--\pageref{lastpage}}
\maketitle

\begin{abstract}
Lyman-$\alpha$ damping wings towards quasars provide a unique probe of reionization because their strength correlates strongly with the global volume-averaged neutral hydrogen (HI) fraction of the intergalactic medium (IGM). Cosmic variance in the IGM, however, is a major source of stochasticity since the \textit{local} neutral environment around a quasar varies significantly even at fixed global neutral fraction. We show that the IGM damping wing carries additional information about this local ionization topology, unexploited by current analysis frameworks. We introduce a set of two new physically motivated summary statistics encoding the local information about the HI distribution in the IGM \textit{before} it is altered by ionization radiation from the quasar, encompassing 1) the HI column density, weighted by a Lorentzian profile mimicking the frequency dependence of the Lyman-$\alpha$ cross section, and 2) the distance from the quasar to the first neutral patch. This description, when combined with the quasar's lifetime as a third parameter, reduces the IGM transmission scatter in the damping wing region of the spectrum to $\lesssim 1\,\%$ across the full range of physical parameter space. We introduce a simple procedure for generating synthetic HI sightlines around quasars and demonstrate that the resulting damping wing profiles are statistically indistinguishable from a realistic reionization topology. This opens the door for optimally extracting the salient local information encoded in the imprint in a model-independent fashion. In the context of a specific reionization model, measurements of these local parameters can be translated into constraints on the global timing of reionization, but in addition, they provide information about the reionization topology, hitherto unused. A marginally modified version of our framework can also be employed in the context of damping wings towards galaxies.

\end{abstract}

\begin{keywords}
intergalactic medium -- quasars: absorption lines -- cosmology: observations -- cosmology:
theory -- dark ages, reionization, first stars.
\end{keywords}

\section{Introduction}

The epoch of reionization was a hallmark event in the evolution of our universe when the until then prevailing dark ages were ended by the light from the first galaxies and quasars. It is these very objects which we can also use as highly sensitive torchlights probing the presence of even the smallest amounts of neutral hydrogen in the intergalactic medium (IGM). This manifests in the form of extended Gunn-Peterson absorption troughs \citep{gunn1965} appearing at global volume-averaged IGM neutral fractions as low as $\langle x_\mathrm{HI} \rangle \gtrsim 10^{-4}$ where the Lyman-$\alpha$ transition saturates, and transitions into a smooth damping wing imprint redward of the Lyman-$\alpha$ line when $\langle x_\mathrm{HI} \rangle$ reaches order unity \citep{miralda-escude1998}. As such, every single high-redshift source offers a glimpse at the progress of reionization
of the universe at its respective redshift. Existing \citep{dodorico2023, onorato2024} and future \citep{euclid_collaboration2019} statistical ensembles of their spectra provide us with a unique avenue to constrain the full timeline of the epoch of reionization, a notoriously hard problem due to the model-dependence
of all probes going beyond the CMB electron scattering optical depth \citep{planck_collaboration2020} and Lyman-series dark pixel fractions \citep[e.g.][]{mcgreer2015, jin2023}.

In the case of IGM damping wings, this undertaking is exacerbated by the fact that the IGM absorption imprint has to be disentangled from the intrinsic spectra of these sources, and, potentially, local damped Lyman-$\alpha$ (DLA) absorption systems in front of the source which can mimic the cosmological imprint from the IGM \citep{huberty2025}. While both these tasks remain tractable when considering quasars as background sources, additional complications arise due to the patchy nature of the reionization process. A consequence of this is that at fixed IGM neutral fraction $\langle x_\mathrm{HI} \rangle$, the column density of neutral material
\textit{along a given sightline} can vary drastically, depending on how many neutral patches it actually contains. The distribution of neutral versus ionized sightlines at given $\langle x_\mathrm{HI} \rangle$ depends on the topology 
of reionization. The relation between damping wing strength and global IGM neutral fraction is thus inevitably \textit{stochastic}. In essence, the Lyman-$\alpha$ damping wing only probes the \textit{local} ionization state of the surrounding IGM as a single, statistical draw from the \textit{global} ionization topology at the corresponding redshift.

To date, existing damping wing analysis frameworks \citep[e.g.][]{greig2017b, greig2024, davies2018a, durovcikova2020, durovcikova2024, hennawi2024, umeda2024, umeda2025, mason2025} have not made a clear methodological distinction between the two. 
The conventional approach has always consisted of directly inferring (among other parameters) the \textit{global} IGM neutral fraction $\langle x_\mathrm{HI} \rangle$, %
building upon analytical expressions for the damping wing optical depth \citep{miralda-escude1998}, or realistic IGM transmission profiles based on cosmological simulations. As demonstrated recently in the context of \textit{quasar} damping wings \citep{kist2025}, not less than half the total error budget on the inferred IGM neutral fraction $\langle x_\mathrm{HI} \rangle$ arises due to density fluctuations and the stochasticity of reionization, comparable to the uncertainty due to the reconstruction of the unknown quasar continuum. An additional major source of uncertainty in the context of \textit{galaxy} damping wings are local proximate Lyman-$\alpha$ absorption systems \citep{heintz2024, heintz2025} which can obfuscate the imprint arising from the IGM \citep{huberty2025}, and would need to be marginalized out in order to obtain faithful constraints on the ionization state of the IGM \citep{mason2025}.

In any case, eliminating the uncertainty sourced by the stochasticity of reionization would allow us to extract \textit{all} the information that the damping wing imprint actually encodes---that is, physical line-of-sight information about the neutral hydrogen (HI) density field in front of the quasar which is informative not only about the timing but also the topology of reionization. Instead of directly processing
this information into a constraint on the \textit{global} IGM neutral fraction, leaving the latter unconstrained, we here propose a novel set of \textit{physical} summary statistics which quantify the \textit{local} HI content in front of a given object, and hence tightly parameterize the characteristic shape of the IGM damping wing.

Finding a better parameterization of the IGM damping wing has also been the primary motivation of \citet{chen2024} who proposed to re-center IGM transmission profiles based on the position in the spectrum where the transmission first approaches zero, and \citet{keating2024a} who showed that the damping wing strength is tightly parameterized by the average HI density weighted with a Lorentzian profile, mimicking the Lorentzian decay of the Lyman-$\alpha$ cross section. Furthermore, \citet{mason2025} for the first time inferred the distance between the source and the first neutral bubble as a local measure in addition to the global IGM neutral fraction $\langle x_\mathrm{HI} \rangle$. Here we
build upon the labels introduced by \citet{keating2024a} and \citet{mason2025}, and establish two clearly defined physical summary statistics quantifying the local HI content around the source, readily applicable for astrophysical parameter inference. Leveraging these summary statistics, we show that the damping wing imprint does not only bear information about the \textit{timing} of reionization in form of the evolution of the global IGM neutral fraction as a function of redshift, but even \textit{topological} information about, e.g., the distribution of ionized bubble sizes around the most massive halos \textit{before quasars started shining}.

In addition to that, we show that our summary statistics are statistically insensitive to the underlying reionization model. Instead, all stochasticity in this parameterization is sourced by the well-understood distribution of density fluctuations in the IGM. This paves the way for \textit{topology-independent}, local damping wing constraints which carry topological information, and can be tied \textit{subsequently} to a specific reionization topology, constraining, e.g., the global IGM neutral fraction $\langle x_\mathrm{HI} \rangle$ \textit{in the context of that given topology}.

We start by introducing our formalism and defining our new local summary statistics in Section~\ref{sec:labels}. We quantify the scatter of realistic IGM transmission profiles in the damping wing region within this parameterization in Section~\ref{sec:real_top} and proceed in Section~\ref{sec:topology} by introducing a simplistic toy reionization bubble model which we then use to demonstrate the topology-independence of our parameterization. We conclude in Section~\ref{sec:conclusions}.

\section{Towards a three-parameter model of quasar IGM damping wings}
\label{sec:labels}

\begin{figure*}
	\includegraphics[width=\textwidth]{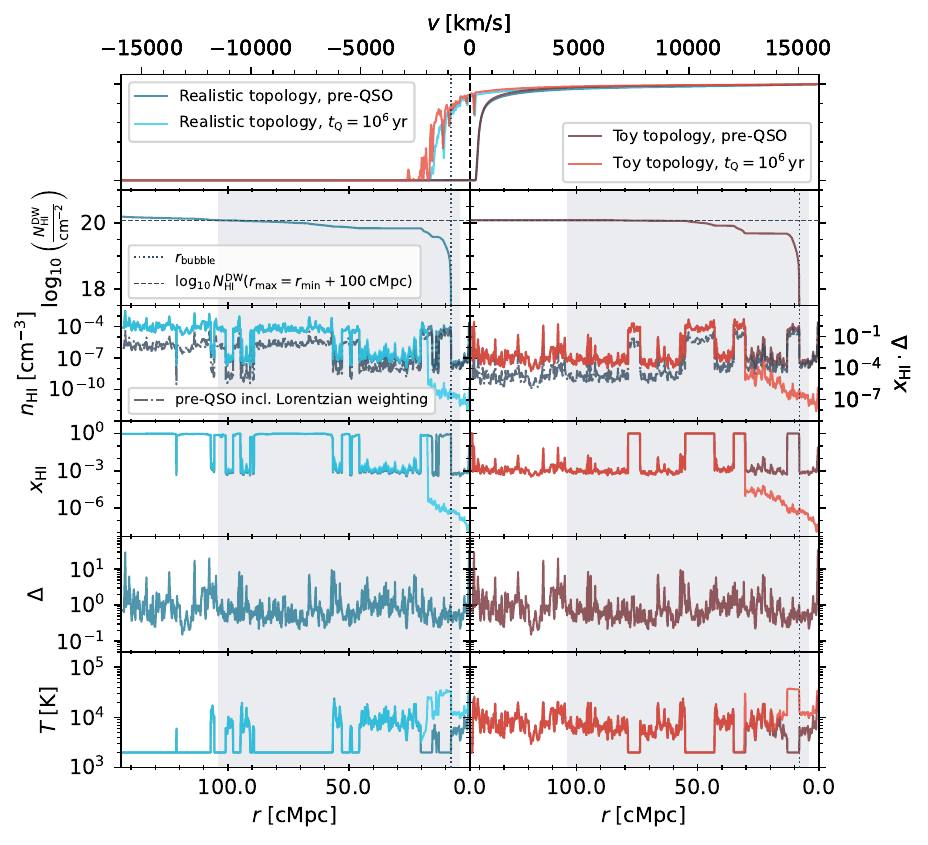}
    \caption{Lyman-$\alpha$ transmission and physical fields for an example Nyx sightline combined with a semi-numerical $x_\mathrm{HI}$ skewer extracted from a $\langle x_\mathrm{HI} \rangle = 0.65$ topology (left panels/blue lines; see Section~\ref{sec:sims}) and an analytical $x_\mathrm{HI}$ skewer generated according to our toy bubble prescription (right panels/red lines; see Section~\ref{sec:bubble_model}), before (dark) 
    and after (light colors) a quasar has been shining for $t_\mathrm{Q} = 10^6\,\mathrm{yr}$. The panels show from top to bottom the Lyman-$\alpha$ transmission field, the distance-weighted HI column density $N_\mathrm{HI}^\mathrm{DW}$ as a function of the upper integration limit $r_\mathrm{max}$, the HI density field $n_\mathrm{HI}$ (also shown in units of $x_\mathrm{HI} \cdot \Delta$), including its distance-weighted pre-quasar version (black dash-dotted line), the HI fraction $x_\mathrm{HI}$, the overdensity field $\Delta$, and the temperature field $T$. Our fiducial integration range for the distance-weighted HI column density $N_\mathrm{HI}^\mathrm{DW}$ is highlighted by the grey-shaded region, and the horizontal dashed line marks the value of $\log_{10}N_\mathrm{HI}^\mathrm{DW}/\mathrm{cm}^{-2} = 20.1$ corresponding to this integration range. The vertical dotted lines mark the location $r_\mathrm{bubble} = 8\,\mathrm{cMpc}$ of the first neutral bubble in the pre-quasar sightline.}
    \label{fig:skewers}
\end{figure*}

Several works have recently been pushing towards a parameterization better capturing the characteristic shape of IGM damping wings \citep{chen2024, keating2024a}. Specifically, \citet{keating2024a} noted that (in the absence of a strong ionizing source) the distance-weighted average HI number density calculated along simulated sightlines from the Sherwood-Relics simulation suite \citep{bolton2017, puchwein2023} coincides remarkably well with predictions from the analytical \citet{miralda-escude1998} damping wing model. Here, we build upon that finding and introduce novel two-parameter summary statistics of the local \textit{pre}-quasar HI field that significantly reduce the scatter of the damping wing optical depth $\tau_\mathrm{DW}$ as compared to the global IGM neutral fraction $\langle x_\mathrm{HI} \rangle$.

\subsection{Introducing a local HI column density label}
\label{sec:NHI_DW}

We start by recalling that the damping wing optical depth $\tau_\mathrm{DW}$ of a strong ionizing source 
at redshift $z_\mathrm{QSO}$ as a function of rest-frame wavelength $\lambda_\mathrm{rest}$ is obtained by integrating the product of the HI density field $n_\mathrm{HI}^\mathrm{post}$ and the Lyman-$\alpha$ cross section of neutral hydrogen $\sigma_\alpha$
(evaluated at wavelength $\lambda = \lambda_\mathrm{rest}\,(1+z_\mathrm{QSO})/(1+z)$) along the (proper) line-of-sight interval $\mathrm{d}R$:
\begin{equation}
\label{eq:tau_DW}
    \tau_\mathrm{DW}(\lambda_\mathrm{rest}) = \int_{0}^{R(z_\mathrm{QSO})} n_\mathrm{HI}^\mathrm{post}(R) \cdot \sigma_\alpha\left(\frac{1+z_\mathrm{QSO}}{1+z(R)}\,\lambda_\mathrm{rest}\right) \; \mathrm{d}R,
\end{equation}
where $R(z_\mathrm{QSO})$ is the proper distance from the observer to the source, and $z(R)$ is the inverted distance-redshift relation, and we henceforth denote all associated comoving distances with a small $r$ in place of a capital one.

Our goal is to construct summary statistics of the HI density field that best capture the characteristic shape of the IGM damping wing as given by $\tau_\mathrm{DW}(\lambda_\mathrm{rest})$. However, note that we are interested in maximizing the information content with respect to the \textit{pre}-quasar HI density field which we denote as $n_\mathrm{HI}^\mathrm{pre}$ (in contrast to the post-quasar field $n_\mathrm{HI}^\mathrm{post}$ giving rise to the actually observed IGM damping wing as per Eq.~(\ref{eq:tau_DW})). This is because $n_\mathrm{HI}^\mathrm{pre}$ directly relates to the \textit{cosmological} reionization topology, removing the impact of the quasar as a local \textit{astrophysical} source of ionizing radiation. In other words, we can decompose the pre-quasar HI field as
\begin{equation}
\label{eq:nHI_gal}
    n_\mathrm{HI}^\mathrm{pre} = \langle n_{\mathrm{H}} \rangle(z_\mathrm{QSO}) \cdot x_\mathrm{HI} \cdot \Delta,
\end{equation}
where $\langle n_{\mathrm{H}} \rangle(z_\mathrm{QSO})$ is the cosmic mean hydrogen density at the redshift of interest, $\Delta$ is the dimensionless matter overdensity field, and $x_\mathrm{HI}$ is the HI fraction field which informs us about the \textit{global} volume-averaged IGM neutral faction $\langle x_\mathrm{HI} \rangle$.\footnote{For brevity, we write $\langle x_\mathrm{HI} \rangle \equiv \langle x_\mathrm{HI} \rangle(z_\mathrm{QSO})$ throughout this work.}

We now aim to condense the information contained in the \textit{field} $n_\mathrm{HI}^\mathrm{pre}$ into a low-dimensional set of physically motivated summary statistics. The canonical choice for such a summary statistic is the HI column density $N_\mathrm{HI} = \int n_\mathrm{HI}^\mathrm{pre}(R) \; \mathrm{d}R$ along the line of sight. Noting the obvious similarity between this definition and Eq.~(\ref{eq:tau_DW}), and following the weighting proposed in \citet{keating2024a}, we introduce a weighting function
\begin{equation}
\label{eq:weighting}
    w \equiv (R / R_\mathrm{T} + 1)^{-2} \sim (v - v_\mathrm{T})^{-2},
\end{equation}
mimicking the Lorentzian decay of the Lyman-$\alpha$ cross section $\sigma_\alpha$ in Eq.~(\ref{eq:tau_DW}), evaluated at a fixed reference
distance $R = R_\mathrm{T}$.\footnote{Note that away from the Lyman-$\alpha$ resonance, a Lorentzian profile is an excellent approximation of the full quantum-mechanical Lyman-$\alpha$ cross section $\sigma_\alpha$.} The equivalence with a Lorentzian profile becomes more apparent if we introduce the velocity grid $v = - H(z_\mathrm{QSO}) \, R$, centered at the Lyman-$\alpha$ line (such that $\lambda_\mathrm{rest} = \lambda_\alpha\,(1+\frac{v}{c})$ for the Lyman-$\alpha$ wavelength $\lambda_\alpha \simeq 1215.67\,\text{\AA}$), and express $R_\mathrm{T}$ as a velocity offset $v_\mathrm{T} \equiv +H(z_\mathrm{QSO})\,R_\mathrm{T}$.\footnote{Note that our velocity grid follows the standard convention of $v > 0$ redward of the Lyman-$\alpha$ line, and $v < 0$ blueward where $R > 0$. However, to keep all distances positive‚ we define $R_\mathrm{T} > 0$ even though it corresponds to a \textit{positive} velocity offset $v_\mathrm{T} > 0$ \textit{redward} of Lyman-$\alpha$.} 

Given a field $X$, we now define the line-of-sight average $\llangle . \rrangle_{\mathrm{Lor}}$ of this field with respect to the Lorentzian weighting kernel defined in Eq.~(\ref{eq:weighting}) as
\begin{equation}
\label{eq:Lorentzian_avg}
    \llangle X \rrangle_{\mathrm{Lor}} \equiv \frac{1}{\mathcal{N}(u_\mathrm{min}, u_\mathrm{max})} \, \int_{u_\mathrm{min}}^{u_\mathrm{max}}  \frac{X(u)}{(u + 1)^2} \; \mathrm{d}u,
\end{equation}
where we introduced the dimensionless integration variable $u\equiv R/R_\mathrm{T}$ with integration limits $u_\mathrm{min}$ and $u_\mathrm{max}$, corresponding to the distances $R_\mathrm{min}$ and $R_\mathrm{max}$ whose values we will fix below. Further, $\mathcal{N}(u_\mathrm{min}, u_\mathrm{max})$ is a normalization factor ensuring $\llangle \boldsymbol{1} \rrangle_{\mathrm{Lor}} = 1$ for the identity field $\boldsymbol{1}$. This factor can be explicitly determined, yielding
\begin{equation}
    \mathcal{N}(u_\mathrm{min}, u_\mathrm{max}) = \frac{u_\mathrm{max}-u_\mathrm{min}}{(u_\mathrm{max}+1)(u_\mathrm{min}+1)}.
\end{equation}

Comparing Eq.~(\ref{eq:Lorentzian_avg}) to Eq.~(\ref{eq:tau_DW}) governing the damping wing optical depth, we now proceed by defining the \textit{distance-weighted}
HI column density as
\begin{equation}
\label{eq:NHI_DW}
    N_\mathrm{HI}^\mathrm{DW} \equiv \mathcal{N}\left(\tfrac{R_\mathrm{min}}{R_\mathrm{T}}, \tfrac{R_\mathrm{max}}{R_\mathrm{T}}\right) \cdot R_\mathrm{T} \cdot \langle n_\mathrm{H} \rangle (z_\mathrm{QSO})  \cdot  \llangle x_\mathrm{HI} \cdot \Delta \rrangle_{\mathrm{Lor}}.
\end{equation}
This definition involves three hyperparameters, i.e., the reference distance $R_\mathrm{T}$ as well as the integration limits $R_\mathrm{min}$ and $R_\mathrm{max}$. An informed choice of these parameters ensures that this summary statistic reflects the characteristic shape of the IGM damping wing not only for the pre-quasar topology (which we would like to maximize the information content about), but also the post-quasar one (which, due to the impact of the quasar ionizing radiation, inevitably carries a somewhat decreased amount of information about the pre-quasar IGM, but on the other hand gives rise to the actually observed damping wing imprint as per Eq.~(\ref{eq:tau_DW})).

Firstly, with regards to the \textit{pre}-quasar topology, we show in Appendix~\ref{app:tau_vs_NHI} that in the limit where $R_\mathrm{min} \to 0$ and $R_\mathrm{max} \to R(z_\mathrm{QSO})$, and where the Lyman-$\alpha$ cross section can be approximated as perfectly Lorentzian, the pre-quasar optical depth $\tau_\mathrm{DW}^\mathrm{pre}$ (as defined in Eq.~(\ref{eq:tau_DW_pre})) evaluated at the velocity offset $v_\mathrm{T}$ becomes, up to a prefactor, approximately proportional to $N_\mathrm{HI}^\mathrm{DW}$:
\begin{equation}
\label{eq:tau_DW_approx}
    \tau_\mathrm{DW}^\mathrm{pre}(v=v_\mathrm{T}) \simeq \frac{e^2}{m_e c^2}\,f_\alpha\,\gamma_\alpha\,\lambda_\alpha\,(c/v_\mathrm{T}-1)^2 \times N_\mathrm{HI}^\mathrm{DW},
\end{equation}
where $f_\alpha\simeq 0.416$ is the Lyman-$\alpha$ oscillator strength, and $\gamma_\alpha \equiv \Gamma_\alpha\,\lambda_\alpha/(4\pi c)$ with the Lyman-$\alpha$ decay constant $\Gamma_\alpha=6.265\,\times\,10^8\,\mathrm{s}^{-1}$.
This motivates our choice of $N_\mathrm{HI}^\mathrm{DW}$ as the primary summary statistic that minimizes the scatter of the red-side damping wing transmission among different (pre-quasar) sightlines, and hence encodes the characteristic shape of the IGM damping wing in a single number.

Practically, however, we are concerned with profiles originating from the \textit{post}-quasar topology rather than the pre-quasar one. As we will show in the consequent sections, $r_\mathrm{min} = 4\,\mathrm{cMpc}$, $r_\mathrm{max} = r_\mathrm{min} + 100\,\mathrm{cMpc}$, and $r_\mathrm{T} = 18\,\mathrm{cMpc}$ (corresponding to $v_\mathrm{T} \simeq 2000\,\mathrm{km}/\mathrm{s}$ at $z_\mathrm{QSO} = 7.54$)\footnote{As our analysis is restricted to models at this fixed redshift, we henceforth interchangeably refer to reference distances $r_\mathrm{T}$ and reference velocity offsets $v_\mathrm{T}$, noting that only a fixed choice of $r_\mathrm{T}$ would account for redshift evolution in the density field.} are adequate choices to ensure that Eq.~(\ref{eq:tau_DW_approx}) remains valid even for the post-quasar optical depth $\tau_\mathrm{DW}$. Adopting these parameter choices and converting all quantities to comoving units, we can evaluate Eq.~(\ref{eq:NHI_DW}) as
\begin{align}
\label{eq:NHI_DW_vals}
    N_\mathrm{HI}^\mathrm{DW} =\; &5.1 \times 10^{20}\,\mathrm{cm}^{-2}\times \left(\frac{\mathcal{N}\left(\tfrac{r_\mathrm{min}}{r_\mathrm{T}}, \tfrac{r_\mathrm{max}}{r_\mathrm{T}}\right)}{0.67}\right) \left(\frac{r_\mathrm{T}}{18\,\mathrm{cMpc}}\right) \\ \nonumber
    &\times \left(\frac{1+z_\mathrm{QSO}}{1+7.54}\right)^2 \left(\frac{\llangle x_\mathrm{HI} \cdot \Delta \rrangle_{\mathrm{Lor}}}{1}\right).
\end{align}
Here we used $\langle n_\mathrm{H} \rangle(z_\mathrm{QSO}) = \langle n_\mathrm{H} \rangle_{{}_0}(1+z_\mathrm{QSO})^3$, where $\langle n_\mathrm{H} \rangle_{{}_0} = 3H_0^2(1-Y)\Omega_{\rm b}/(8\pi G m_{\rm p}) \simeq 1.9\times10^{-7} \; {\rm atoms}/{\rm cm}^3$ with $H_0$ and $\Omega_{\rm b}$ chosen according to a \citet{planck_collaboration2020} cosmology, and a big bang nucleosynthesis (BBN) hydrogen fraction of $1-Y = 0.76$.

We show in the top and left panels of Figure~\ref{fig:skewers} how we arrive at this summary statistic for a simulated example sightline of a $t_\mathrm{Q} = 10^6\,\mathrm{yr}$ quasar in a globally $65\,\%$ neutral universe. A more detailed description of the underlying simulations will be provided in Section~\ref{sec:sims}. In each panel, the post-quasar version of the corresponding field is depicted in light blue, the pre-quasar one in dark blue. The top panel shows the IGM transmission profiles,\footnote{Note that all profiles which are colored red will be discussed later in Section~\ref{sec:topology}.} and the second panel shows the distance-weighted HI column density (Eq.~(\ref{eq:NHI_DW_vals})) as a function of the upper integration limit $r_\mathrm{max}$, where the $N_\mathrm{HI}^\mathrm{DW}$ value resulting from our fiducial choice of $r_\mathrm{max} = r_\mathrm{min} + 100\,\mathrm{cMpc}$ is marked by the horizontal dotted line. The third panel depicts the HI density field $n_\mathrm{HI}$ (also given in units of $x_\mathrm{HI}\cdot\Delta$), and additionally the distance-weighted version of the pre-quasar field according to Eq.~(\ref{eq:weighting}). The remaining panels show the neutral hydrogen fraction field $x_\mathrm{HI}$, the overdensity field $\Delta$, and the temperature field $T$, respectively. The grey-shaded area in each panel depicts our fiducial integration range of $r_\mathrm{min} = 4\,\mathrm{cMpc}$ and $r_\mathrm{max} = r_\mathrm{min} + 100\,\mathrm{cMpc}$. We can see how distant neutral patches are downweighted by the Lorentzian weighting function (third panel), and therefore contribute less to $N_\mathrm{HI}^\mathrm{DW}$ (second panel) and hence do not notably affect the IGM damping wing (top panel) either.

\subsection{Choice of the $N_\mathrm{HI}^\mathrm{DW}$ integration limits}
\label{sec:int_limits}

As Eq.~(\ref{eq:tau_DW_approx}) shows, our $N_\mathrm{HI}^\mathrm{DW}$ statistic is an optimal prescription of the damping wing optical depth $\tau_\mathrm{DW}^\mathrm{pre}(v=v_\mathrm{T})$ of the pre-quasar density field $n_\mathrm{HI}^\mathrm{pre}$ evaluated at the velocity offset $v_\mathrm{T}$, provided that we integrate along the \textit{entire} line of sight and assume a perfectly Lorentzian cross section. If we were to consider damping wings around galaxies, it would therefore be immediately clear that we would have to set our lower integration limit to $r_\mathrm{min} = 0$.

However, this simple picture changes when considering damping wings towards quasars instead. Even though we are still ultimately interested in the pre-quasar HI field, in practice we are inevitably probing the optical depth arising from the \textit{post}-quasar field $n_\mathrm{HI}^\mathrm{post}$. Differences arise interior to the quasar ionization front where the pre-quasar neutral material is ionized away and hence the information about this part of the pre-quasar density field is lost; though we can account for this by also constraining the lifetime $t_\mathrm{Q}$ of the quasar. By tuning our integration limits $R_\mathrm{min}$ and $R_\mathrm{max}$ in Eq.~(\ref{eq:NHI_DW}), we can make sure that we still maximize the information content about $n_\mathrm{HI}^\mathrm{pre}$, and nevertheless obtain a statistic which describes the observed damping wing profile with a minimum amount of scatter. By nature of the required approximations, this scatter will be non-zero, but an informed choice of the integration limits ensures that it stays minimal and certainly remains smaller that the amount of scatter which $\langle x_\mathrm{HI} \rangle$ entails.

To do so, we firstly note that the optical depth integral in Eq.~(\ref{eq:tau_DW}) effectively only starts at the radius of the ionization front $R_\mathrm{ion}$ that the quasar has carved out, since $n_\mathrm{HI}^\mathrm{post}$ is drastically reduced within the ionized bubble around the quasar (i.e., at $R<R_\mathrm{ion}$). On the other hand, at $R>R_\mathrm{ion}$, where the quasar ionization front has not arrived yet, we can safely approximate $n_\mathrm{HI}^\mathrm{post} \simeq n_\mathrm{HI}^\mathrm{pre}$. In a simple theoretical picture where every photon emitted by the quasar ionizes exactly one hydrogen atom in a (Strömgren) sphere around the quasar, the radius of the ionization front $R_\mathrm{ion}$ relates to the lifetime of the quasar $t_\mathrm{Q}$ and the volume averaged HI density $\langle n_\mathrm{HI}^\mathrm{pre} \rangle \equiv \langle n_\mathrm{HI}^\mathrm{pre} \rangle(z_\mathrm{QSO})$ as
\begin{equation}
\label{eq:R_ion}
R_{\rm ion} = \left(\frac{3 Q \, t_{\rm Q}}{4\pi \langle n_\mathrm{HI}^\mathrm{pre} \rangle}\right)^{1/3}
\end{equation}
\citep{cen2000}, where $Q$ is the quasar's emission rate of ionizing photons. Due to the approximate nature of this equation, defining $R_\mathrm{min} \equiv R_\mathrm{ion}$ is not guaranteed to give us the correct integration limit for an individual sightline. However, as we strive to define a sightline-independent integration limit, Eq.~(\ref{eq:R_ion}) constitutes a useful starting point for determining a universal value.

Since we aim to define our $N_\mathrm{HI}^\mathrm{DW}$ label agnostically to the reionization state of the universe, we first have to remove the dependence of the cosmic mean HI density $\langle n_\mathrm{HI}^\mathrm{pre} \rangle = \langle n_\mathrm{H} \rangle_{{}_0} (1+z_\mathrm{QSO})^3 \langle x_\mathrm{HI} \rangle$ on the IGM neutral fraction $\langle x_\mathrm{HI} \rangle$. We can achieve this by assuming $\langle x_\mathrm{HI} \rangle = 1$ which gives rise to the shortest $R_\mathrm{ion}$ possible. By doing so, we make our lower integration limit shorter, and hence conservatively also account for regions closer to the source where neutral material is only still present for sightlines from highly neutral topologies.

Secondly, as $N_\mathrm{HI}^\mathrm{DW}$ is supposed to be a summary statistic for the \textit{pre}-quasar HI density $n_\mathrm{HI}^\mathrm{pre}$, the lifetime dependence of Eq.~(\ref{eq:R_ion}) is undesirable too, and we can eliminate it by fixing $t_\mathrm{Q}$ in Eq.~(\ref{eq:R_ion}) to a representative value of $t_\mathrm{Q} = 10^{4.5}\,\mathrm{yr}$. This specific value constitutes an empirical compromise between 1) quasars with longer lifetimes of $t_\mathrm{Q} \gtrsim 10^{4.5}\,\mathrm{yr}$ and hence larger $R_\mathrm{ion}$, where a fixed lower integration limit of $R_\mathrm{ion}(t_\mathrm{Q} = 10^{4.5}\,\mathrm{yr})$ includes pre-quasar IGM neutral material which has already been ionized away and therefore does \textit{not} contribute to the damping wing, and 2) short-lived quasars with $t_\mathrm{Q} \lesssim 10^{4.5}\,\mathrm{yr}$ where this integration limit excludes foreground neutral material that actually \textit{will} impact the damping wing.
As we will demonstrate in Section~\ref{sec:local_scatter}, our choice of $t_\mathrm{Q} = 10^{4.5}\,\mathrm{yr}$ provides a reasonably accurate approximation for the full lifetime range expected for observed high-redshift quasars \citep[see e.g.][]{morey2021, khrykin2021}. 
With these choices, we arrive at the comoving value of
\begin{equation}
\label{eq:R_ion_num}
    r_{\rm ion} = 4.0\,{\rm cMpc}\left(\frac{Q}{10^{57.14}\,{\rm s^{-1}}}\right)^{1\slash 3}\left(\frac{t_{\rm Q}}{10^{4.5}\,{\rm yr}}\right)^{1\slash 3}\left(\frac{\langle x_{\rm HI}\rangle}{1.0}\right)^{-1\slash 3}
\end{equation}
and fix our lower integration limit accordingly to $r_\mathrm{min} = 4.0\,{\rm cMpc}$.\footnote{Note that due to cosmological expansion manifesting in the density field, we arrived at the same comoving integration limit independently of redshift.}

The optical depth integral (Eq.~(\ref{eq:tau_DW})) formally runs all the way down to redshift zero; however, due to the Lorentzian decline in the outer parts of the HI cross section, we can safely stop the integration in Eq.~(\ref{eq:Lorentzian_avg}) already significantly earlier than this, while still retaining a tight approximation of the full damping wing optical depth as per Eq.~(\ref{eq:tau_DW_approx}). Specifically, we only have to integrate until the Lorentzian line-of-sight average of neutral fraction and overdensity field $\llangle x_\mathrm{HI} \cdot \Delta \rrangle_{\mathrm{Lor}}$ has converged. We practically find that this is the case after integrating over a range of $\sim 100\,\mathrm{cMpc}$ (as exemplarily seen in the second panel of Figure~\ref{fig:skewers}), where our weighting kernel (Eq.~(\ref{eq:weighting})) has decreased to $\sim 3\,\%$ of its value at $r_\mathrm{min}$. Thus, we set the upper integration limit to $r_\mathrm{max} = r_\mathrm{min} + 100\,\mathrm{cMpc}$.

\subsection{The damping wing as a one-parameter family}
\label{sec:one_param_family}

\begin{figure}
	\includegraphics[width=\columnwidth]{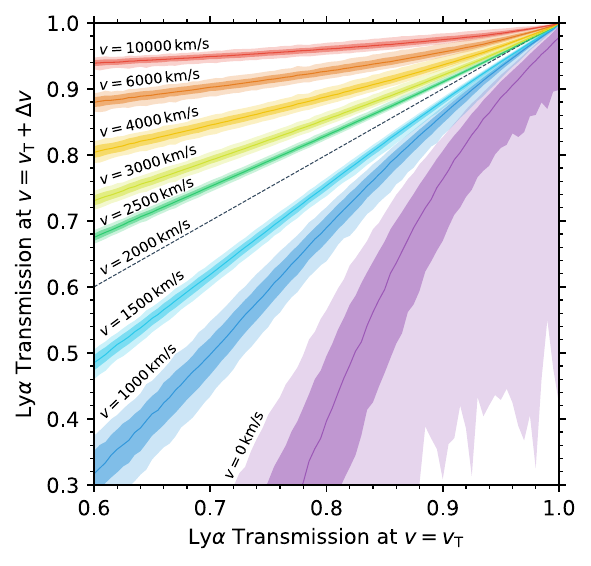}
    \caption{IGM transmission values at the fixed reference velocity offset $v_\mathrm{T} = 2000\,\mathrm{km}/\mathrm{s}$ versus transmission values at a number of velocities offsets $v = v_\mathrm{T} + \Delta v$ away from this reference offset. Solid curves and shaded regions depict the median as well as the $68$- and $95$-percentile scatter of these transmission values, respectively, among $852 \times 21 \times 51$ realistic IGM transmission profiles simulated according to the description in Section~\ref{sec:sims}. For reference, we are also showing the trivial one-to-one relation for $v = v_\mathrm{T}$ as a dashed black line.} %
    \label{fig:transm_corr}
\end{figure}

We argued in Section~\ref{sec:NHI_DW} that we identified a close-to optimal summary statistic $N_\mathrm{HI}^\mathrm{DW}$ for the damping wing optical depth evaluated at a fixed reference distance $r_\mathrm{T}$ (or, equivalently, velocity offset $v_\mathrm{T}$). We now demonstrate that by restricting ourselves to this fixed reference point, we do not give up a significant amount of additional information encoded in the imprint. In other words, we show that the Lyman-$\alpha$ transmission values at different velocity offsets are so correlated with each other that the damping wing profile essentially constitutes a one-parameter family, i.e., the entire spectral shape is determined by a single parameter, namely the transmission $\tau_\mathrm{DW}(v=v_\mathrm{T})$ at our reference velocity offset.

To demonstrate this, we depict in Figure~\ref{fig:transm_corr} the correlation between a large number of simulated IGM transmission profiles evaluated at our fixed reference velocity offset $v_\mathrm{T} = 2000\,\mathrm{km}/\mathrm{s}$ versus their value at a number of other velocity offsets $v = v_\mathrm{T} + \Delta v$. Specifically, we considered $852 \times 21 \times 51$ transmission profiles consisting of $852$ distinct sightlines at $21$ global IGM neutral fraction values $\langle x_\mathrm{HI} \rangle$ covering a linear range between $0$ and $1$, and $51$ lifetime values $t_\mathrm{Q}$ covering a logarithmic range between $10^3$ and $10^8\,\mathrm{yr}$. The technical details of the simulations will be outlined in Section~\ref{sec:sims}.

For each velocity offset, we show the median transmission values of all profiles as a solid curve, as well as the $68$- and $95$-percentile scatter as shaded regions. Our first observation is that in the largest parts of the damping wing region (at all velocity offsets $v\gtrsim 1500\,\mathrm{km}/\mathrm{s}$), the width of the $68$- ($95$-) percentile region never exceeds $\sim 1 - 2\,\%$ ($\sim 2 - 3 \,\%$), even at the lowest IGM transmission values which are significantly impacted by damping wing absorption, decreasing down to $< 1\,\%$ ($< 2\,\%$) at higher transmission values. A number of effects conspire such that we observe a relatively similar degree of scatter for most of these curves: due to continuity reasons, the scatter naturally decreases for velocity offsets closer to our reference offset at $v_\mathrm{T}=2000\,\mathrm{km}/\mathrm{s}$ marked by the dashed black line.

However, the scatter does not change symmetrically around $v_\mathrm{T}$, i.e., we find a more significant increase blueward of $v_\mathrm{T}$ compared to redward. As demonstrated by \citet{keating2024b}, this is due to the impact of infalling gas as well as residual neutral gas in ionized bubbles which increasingly affects the shape of the transmission profiles in the vicinity of the Lyman-$\alpha$ line. These effects are also the reason why we cannot expect the damping wing to remain a one-parameter family at velocity offsets even closer to Lyman-$\alpha$ line center. Specifically, we see that the $68$-percentile scatter has already increased to $\sim 6\,\%$ at $v=1000\,\mathrm{km}/\mathrm{s}$, and grows to $O(10\,\%)$ at Lyman-$\alpha$ line center. The scatter also grows in the other direction (i.e, further redward of $v_\mathrm{T}$), albeit significantly more modestly than blueward. In this direction, the scatter is limited by the fact that transmission values are generally higher, approaching their upper bound of unity, and hence allowing for less scatter overall. As a result, the spread of the curves shows a mild peak around $v=4000\,\mathrm{km}/\mathrm{s}$, and then decreases again for even larger velocity offsets as the transmission values approach unity.

Overall, this demonstrates that the IGM transmission value at the reference velocity of $v_\mathrm{T} = 2000\,\mathrm{km}/\mathrm{s}$ informs us about the \textit{entire} shape of the IGM damping wing at \textit{any} velocity offset $v \gtrsim  1500\,\mathrm{km}/\mathrm{s}$ with remarkable precision, and hence a summary statistic that maximizes the information at this reference location also extracts the bulk of the information encoded in the entire damping wing imprint.

\subsection{A second label: the distance to the first neutral bubble}

As discussed in the previous sections, the characteristic shape of the IGM damping wing is encoded to lowest order in the distance-weighted HI column density $N_\mathrm{HI}^\mathrm{DW}$ as defined in Eq.~(\ref{eq:NHI_DW}). By defining a second summary statistic, we can 1) further minimize the scatter among Lyman-$\alpha$ transmission skewers at a given set of parameter values, and 2) gain additional information about the reionization topology.

An obvious choice to achieve the first objective would be the location $r_\mathrm{ion}$ of the ionization front of the quasar, as this determines precisely where the optical depth integral in Eq.~(\ref{eq:tau_DW}) effectively starts (c.f. the discussion in Section~\ref{sec:int_limits}). However, at a fixed ionizing photon emission rate $Q$, and at a fixed column density $N_\mathrm{HI}^\mathrm{DW}$, the radius of the ionization front $r_\mathrm{ion}$ is, up to a minor contribution due to density fluctuations, largely degenerate with the quasar lifetime $t_\mathrm{Q}$, as Eq.~(\ref{eq:R_ion}) suggests. The lifetime, however, is of direct physical interest and we therefore opt to rather keep $t_\mathrm{Q}$ as a post-quasar label.

By instead turning to the \textit{pre}-quasar equivalent of $r_\mathrm{ion}$, i.e., the distance $r_\mathrm{bubble}$ between the source and the first neutral bubble along the line of sight \textit{in the pre-quasar topology}, we can simultaneously address both aforementioned objectives: while it is obvious that most long-lived quasars have carved out an ionized bubble extending significantly beyond the pre-quasar neutral bubble location, the same is not true for objects whose ionization front $r_\mathrm{ion}$ does not or only marginally exceed $r_\mathrm{bubble}$. The latter can be the case primarily for young quasars, but also older ones where the first neutral bubble is located at a large distance from the quasar itself.
In any case, due to the fixed choice of the integration limit $r_\mathrm{min}$ for $N_\mathrm{HI}^\mathrm{DW}$, it is clear that $r_\mathrm{bubble}$ encapsulates 
additional information about the pre-quasar ionization topology
that is not fully captured by $N_\mathrm{HI}^\mathrm{DW}$. Besides this, $r_\mathrm{bubble}$ is of direct physical interest as a key statistic of the topology of reionization, yielding complementary information to the values estimated from
Lyman-$\alpha$ emission from galaxies \citep{mason2020, hayes2023, umeda2024, witstok2024, torralba-torregrosa2024, lu2024a, lu2024b, nikolic2025}, and has already been introduced by \citet{mason2025} as an additional summary statistic for IGM damping wings around galaxies.

Overall, this provides us with a three-parameter model, parameterizing quasar IGM damping wings as a function of $(N_\mathrm{HI}^\mathrm{DW}, r_\mathrm{bubble}, t_\mathrm{Q})$. 
The two labels $(N_\mathrm{HI}^\mathrm{DW}, r_\mathrm{bubble})$ are summary statistics of the \textit{pre}-quasar topology, and could hence also be employed in the context of IGM damping wings towards galaxies, provided that the integration range of $N_\mathrm{HI}^\mathrm{DW}$ is adjusted accordingly. The third parameter $t_\mathrm{Q}$ is specific to quasars, encapsulating their impact as strong ionizing sources on the IGM transmission field.

\subsection{Relation to previous definitions}

Parameterizations aiming to better capture the characteristic shape of the IGM damping wing have recently been proposed in \citet{chen2024}, \citet{keating2024a}, and \citet{mason2025}. Our $N_\mathrm{HI}^\mathrm{DW}$ statistic introduced in Section~\ref{sec:NHI_DW} particularly builds upon the study by \citet{keating2024a}, whereas the distance $r_\mathrm{bubble}$ to the first neutral patch has already been employed by \citet{mason2025} in the context of galaxy damping wings. In this section we investigate the similarities and differences among and these definitions found in the literature and our own ones.

We chose to adopt as our first summary statistic the distance-weighted HI column density $N_\mathrm{HI}^\mathrm{DW}$ along the line of sight from the source, rather than the average velocity-weighted number density $\llangle n_\mathrm{HI} \rrangle_\mathrm{Lor}$ as introduced in \citet{keating2024a}.\footnote{Note that the symbol $\llangle n_\mathrm{HI} \rrangle_\mathrm{Lor}$ is chosen in accordance with our own notation.} We deem this definition more appropriate for the following reasons: as an inherent line-of-sight quantity, $N_\mathrm{HI}^\mathrm{DW}$ emphasizes that the damping wing signature really only encodes one-dimensional information about the \textit{local HI content in front of the quasar}. Only by combining the information from several such sightlines do we gain statistical information about the global average of the HI density and hence the global reionization topology. Secondly, the use of a line-of-sight quantity more clearly emphasizes the \textit{physical scale} (i.e., $\sim 100\,\mathrm{cMpc}$) impacting the IGM damping wing. In addition, the analogy to the column density of a local proximate damped Lyman-$\alpha$ absorber (DLA) helps building intuition for the physical range of $N_\mathrm{HI}^\mathrm{DW}$, as well as the degree to which IGM damping wings and proximate DLAs can be disentangled.

We emphasize, however, that in practice, the notions of HI column densities and line-of-sight averaged HI number densities are equivalent, and we could easily convert between the two by realizing that $\llangle n_\mathrm{HI} \rrangle_\mathrm{Lor} = \langle n_\mathrm{H} \rangle (z_\mathrm{QSO})  \cdot  \llangle x_\mathrm{HI} \cdot \Delta \rrangle_{\mathrm{Lor}}$ as follows from Eqs.~(\ref{eq:nHI_gal}) and (\ref{eq:Lorentzian_avg}). According to Eq.~(\ref{eq:NHI_DW}), this implies that the two statistics only differ by the geometrical factor $\mathcal{N}\left(\tfrac{r_\mathrm{min}}{r_\mathrm{T}}, \tfrac{r_\mathrm{max}}{r_\mathrm{T}}\right)$ as well as a factor of $r_\mathrm{T}$. However, this equality demands that the average $\llangle n_\mathrm{HI} \rrangle_\mathrm{Lor}$ be computed over the same integration range as $N_\mathrm{HI}^\mathrm{DW}$. The adequate lower integration limit for galaxy damping wings as considered in \citet{keating2024a} would be $r_\mathrm{min} = 0$, in contrast to our choice of $r_\mathrm{min} = 4\,\mathrm{cMpc}$ due to the presence of a strong ionizing source which almost certainly ionizes away all neutral gas within the first few $\mathrm{cMpc}$ from the source. Further, with regards to quasar damping wings, \citet{keating2024a} chose to operate on the post-quasar field, whereas we here opt to use pre-quasar labels with an adjusted integration limit $r_\mathrm{min}$. This way, we directly constrain the pre-quasar field---which is of higher physical interest---at the price of a small (and, as we will show, negligible) additional amount of scatter.

In addition, despite the seeming agreement of our weighting function (Eq.~(\ref{eq:weighting})) with that adopted by \citet{keating2024a}, our summary statistic differs in that we do \textit{not} realign the skewers in the way proposed in that work and in \citet{chen2024}. We always keep the velocity grid centered at the Lyman-$\alpha$ line, whereas \citet{chen2024} and \citet{keating2024a} re-center their velocity grid based on the individual realignment point of each transmission profile, defined as the point where the profile approaches zero for the first time. Note further that the meaning of the realignment point differs among the two works: \citet{chen2024} does not perform radiative transfer but attempts to model quasar lifetime effects through a simplistic model for the quasar ionization front. In return, their realignment point thus corresponds to what is conventionally measured as the \textit{proximity zone size} in the spectrum of a quasar, often used as a summary statistic to gain information about its lifetime. \citet{keating2024a}, on the other hand, models lifetime effects through ionizing radiative transfer and performs the realignment at various but \textit{fixed} lifetime values (as well as a set of pre-quasar skewers). Hence, the pre-quasar realignment point encodes similar information as our $r_\mathrm{bubble}$ statistic, while the post-quasar one largely coincides with the proximity zone size \textit{at a fixed lifetime value}, hence largely encoding the scatter due to density fluctuations in the IGM.

As a result, our three-parameter model makes both types of re-centering obsolete: while \citet{chen2024}'s realignment point is well characterized by the quasar lifetime $t_\mathrm{Q}$ which is a separate parameter of our model, \citet{keating2024a}'s pre-quasar realignment point is largely represented by our $r_\mathrm{bubble}$ statistic. We find a negligible impact from realigning post-quasar skewers at a fixed lifetime which, while removing some excess scatter due to density fluctuations in the IGM, would be largely degenerate with $t_\mathrm{Q}$. The virtue of $t_\mathrm{Q}$ and $r_\mathrm{bubble}$ is that both of these summaries are of direct physical interest, and we thus prefer to use them over abstract realignment points.

Furthermore, the distance $r_\mathrm{bubble}$ between the source and the first neutral bubble has been introduced in \citet{mason2025} as an additional summary statistic in the context of galaxy IGM damping wings. Here we adopt the same \textit{pre}-quasar statistic, and show that it remains a meaningful summary even for \textit{post}-quasar IGM transmission profiles.

\section{Quantifying the IGM transmission scatter}
\label{sec:real_top}

\begin{figure*}
	\includegraphics[width=\textwidth]{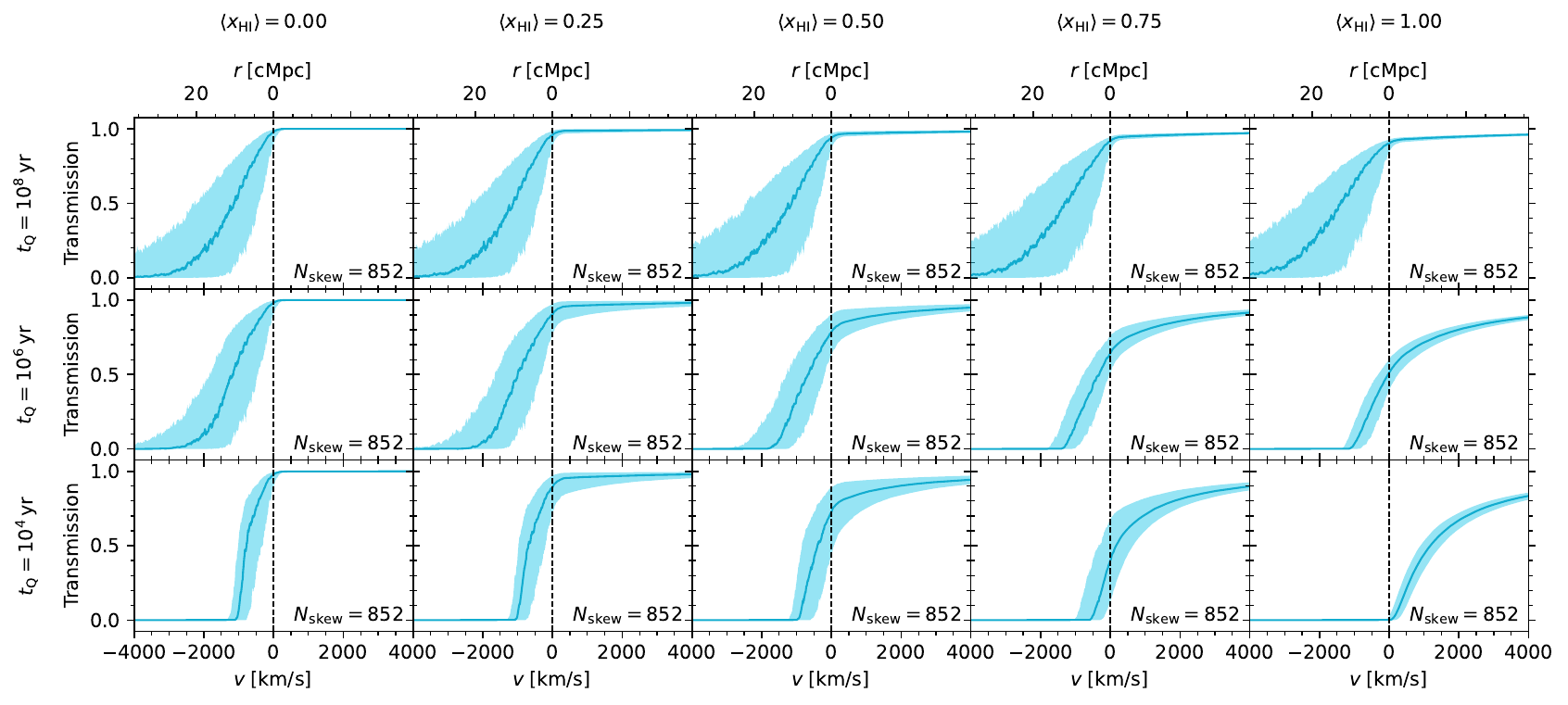}
    \caption{Median and $68$-percentile scatter of the simulated IGM transmission profiles based on our semi-numerical reionization topology in the global $(\langle x_\mathrm{HI} \rangle, t_\mathrm{Q})$ parameterization. The profiles are shown on a representative grid of $(\langle x_\mathrm{HI} \rangle, t_\mathrm{Q})$ parameter values.}
    \label{fig:transm_scatter_xHI}
\end{figure*}

\begin{figure*}
	\includegraphics[width=\textwidth]{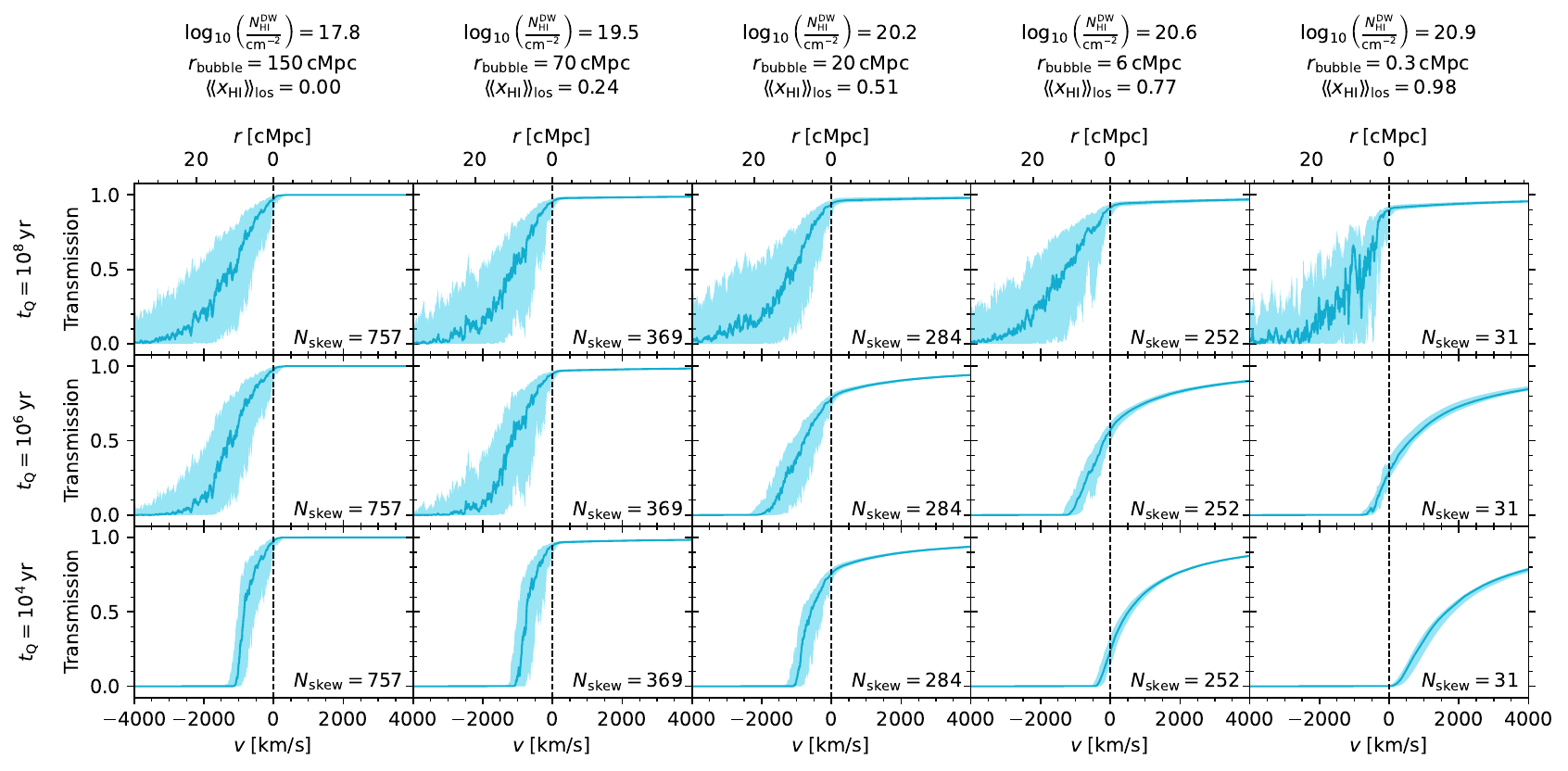}
    \caption{Median and $68$-percentile scatter of the simulated IGM transmission profiles based on our semi-numerical reionization topology in the local $(N_\mathrm{HI}^\mathrm{DW}, r_\mathrm{bubble}, t_\mathrm{Q})$ parameterization. The profiles are shown on a representative grid of $(N_\mathrm{HI}^\mathrm{DW}, r_\mathrm{bubble}, t_\mathrm{Q})$ parameter values for the most typical combinations of $N_\mathrm{HI}^\mathrm{DW}$ and $r_\mathrm{bubble}$ across the full parameter space.}
    \label{fig:transm_scatter_NHI}
\end{figure*}

With the definition of our new set of summary statistics at hand, we now describe how we simulate realistic IGM transmission profiles based on cosmological simulations combined with semi-numerical reionization topologies and 1d radiative transfer. We then use these simulated profiles to demonstrate the advantages 
of our labels over the global IGM neutral fraction $\langle x_\mathrm{HI} \rangle$ in parameterizing the characteristic shape of the IGM damping wing by comparing the IGM transmission scatter at fixed parameter values within the global and the local parameterization.

\subsection{Simulating IGM transmission fields}
\label{sec:sims}

We generate IGM transmission profiles $\boldsymbol{t}$ based on the hybrid approach introduced in \citet{davies2018a}, combining hydrodynamical skewers extracted from the Nyx cosmological simulations and $x_\mathrm{HI}$ skewers from semi-numerical reionization topologies with 1d ionizing radiative transfer.

We start by extracting $6 \times 200$ density, velocity, and temperature skewers originating at the $200$ most massive halos ($M_\mathrm{halo} \geq 2\times 10^{11}\,M_\odot$) from the $z = 7.0$ snapshot of the Nyx simulations \citep{almgren2013, lukic2015}, containing $4096^3$ baryon and another $4096^3$ dark matter particles in a $100\;\mathrm{cMpc}/h$ box. Our fiducial reionization topologies are simulated using an adapted version of \texttt{21cmFast} \citep{mesinger2011, davies2022} in a larger $400\;\mathrm{cMpc}$ box on a $2048^3$ initial and a $512^3$ output grid, providing us with sufficient statistics to probe the distribution of neutral/ionized regions around the $500$ rarest, most massive halos ($M_\mathrm{halo} \geq 3\times 10^{11}\,M_\odot$). We generate such topologies at $21$ different global IGM neutral fractions $\langle x_\mathrm{HI} \rangle$, where we achieve $\langle x_\mathrm{HI} \rangle = 0.05, 0.1, ..., 0.95$ by tuning the ionizing efficiency $\zeta$, and trivially add a completely ionized ($\langle x_\mathrm{HI} \rangle = 0.0$) and a completely neutral ($\langle x_\mathrm{HI} \rangle = 1.0$) model.

We subsequently combine each Nyx sightline with a random $x_\mathrm{HI}$ skewer pointing in a random direction originating at one of the aforementioned $500$ most massive halos of the \texttt{21cmFast} box. Here we adopt the Nyx temperature field for all initially ionized regions, and assume an initially cold IGM for all neutral regions, fixed to $T = 2000\,\mathrm{K}$. To model the small-scale impact of the ionizing quasar radiation, we then perform one-dimensional radiative transfer along these sightlines \citep{davies2016}, assuming a 'light bulb' lightcurve where the quasar has been shining at constant luminosity throughout its lifetime $t_\mathrm{Q}$ until the redshift $z_\mathrm{QSO}$ of interest. We here consider a model with $z_\mathrm{QSO} = 7.54$ and an ionizing photon emission rate of $Q = 10^{57.14}\,\mathrm{s}^{-1}$, resembling the quasar ULAS J1342+0928. We compute the time evolution up to a maximum lifetime of $t_\mathrm{Q} = 10^8\,\mathrm{yr}$, and store $51$ intermediate outputs on a logarithmically spaced grid between $t_\mathrm{Q} = 10^3$ and $10^8\,\mathrm{yr}$. We then convolve the resulting output fields from the radiative transfer code with a Voigt profile to obtain the final IGM transmission profiles. This provides us with a set of $1200$ IGM transmission profiles on a $21 \times 51$ grid of $(\langle x_\mathrm{HI} \rangle, t_\mathrm{Q})$ values. For the bulk
of this work, we focus on the three representative lifetime values of $t_\mathrm{Q} = 10^4$, $10^6$ and $10^8\,\mathrm{yr}$, representing a young, an intermediate, and a long-lived quasar.

We exclude sightlines that exhibit strong proximate optically thick absorption line systems.
This is necessary due to the lack of a subgrid prescription for star formation in the Nyx simulations which implies that the properties and abundance of such systems are not guaranteed to match those found in the real universe.
Observationally, such objects can be excluded based on the presence of associated metal absorption lines in the spectrum \citep{davies2023a}.

To identify such sightlines in our simulations, we consider all $1200$ skewers from the completely ionized pre-quasar topology, divide each of them
into chunks of size $0.1\,\mathrm{pMpc}$, and compute the (unweighted) HI column density within each of these chunks. We then exclude all sightlines where at least one chunk (located within the first $5000\,\mathrm{km}/\mathrm{s}$ from the source) exceeds a column density threshold of $10^{19}\,\mathrm{cm}^{-2}$. 
After applying this criterion, $852$ skewers remain, and we end up with a set of $852 \times 21 \times 51$ transmission profiles covering the full $(\langle x_\mathrm{HI} \rangle, t_\mathrm{Q})$-parameter space.

\subsection{Global parameterization}
\label{sec:global_scatter}

\begin{figure}
	\includegraphics[width=\columnwidth]{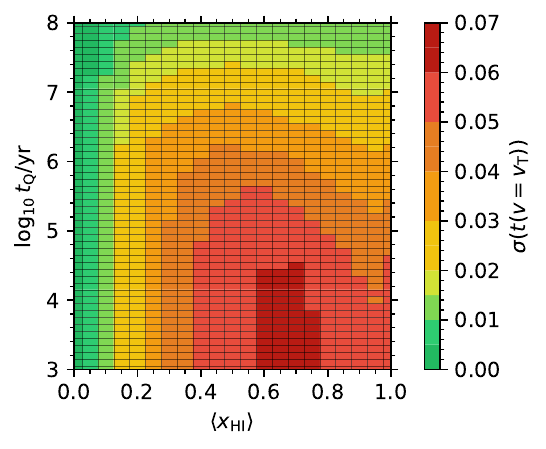}
    \caption{Sightline-to-sightline scatter of the IGM damping wing as a function of astrophysical parameter space in the global $(\langle x_\mathrm{HI}\rangle, t_\mathrm{Q})$-parameterization. The scatter is quantified as half the $68$-percentile width $\sigma(t(v = v_\mathrm{T}))$ of the distribution of IGM transmission values at $v = v_\mathrm{T}$ among all sightlines at a given location in $(\langle x_\mathrm{HI}\rangle, t_\mathrm{Q})$-parameter space.}
    \label{fig:transm_scatter_global_heatmap}
\end{figure}

We start by investigating the IGM transmission scatter within the global parameterization, depicted in Figure~\ref{fig:transm_scatter_xHI} at select $(\langle x_\mathrm{HI} \rangle, t_\mathrm{Q})$ parameter values. Here, the solid lines mark the median transmission value of the $852$ profiles, and the shaded regions denote the $68$-percentile scatter (enclosed by the $16$-th and the $84$-th percentile).
Besides the clear emergence of the damping wing imprint in more neutral environments (i.e., at higher global IGM neutral fractions $\langle x_\mathrm{HI} \rangle$ and shorter quasar lifetimes $t_\mathrm{Q}$), we can see that the bulk of the scatter emerges in the proximity zone due to density fluctuations in the quasar ionized IGM, as well as variation in the location of the quasar ionization front. However, even in the smooth damping wing region redward of the Lyman-$\alpha$ line we can identify a significant amount of scatter.
This scatter is a direct consequence of the stochastic nature of reionization. In other words, at a given global value of the IGM neutral fraction $\langle x_\mathrm{HI} \rangle$, the \textit{local} HI column
density giving rise to the observed damping wing imprint can differ significantly from the global average.

To get a more quantitative idea of this scatter, we focus on the distribution of IGM transmission values at a specific location of the profile, at $v=v_\mathrm{T}$.\footnote{As demonstrated in Section~\ref{sec:one_param_family}, due to the strong correlation across the entire IGM damping wing imprint, it suffices to evaluate $\tau_\mathrm{DW}$ at a single pixel to get a sufficient picture of the scatter across the entire IGM transmission profile.}
Figure~\ref{fig:transm_scatter_global_heatmap} shows a heatmap of the $68$-percentile widths $\sigma(t(v = v_\mathrm{T}))$
of these distributions as a function of the astrophysical parameters $\langle x_\mathrm{HI} \rangle$ and $t_\mathrm{Q}$, where $\sigma(t(v = v_\mathrm{T}))$ is defined as half the distance between $84$-th and $16$-th percentile.

For short quasar lifetimes $t_\mathrm{Q} \lesssim 10^4\,\mathrm{yr}$, we find
similarly wide distributions as \citet{keating2024a} do in profiles from the Sherwood-Relics simulations in the complete absence of a quasar. As expected, 
the scatter decreases with increasing quasar lifetime due to the larger quasar ionized regions
which suppress the IGM damping wing such that the transmission is very close to unity redward of line center. Similarly, the scatter is smaller than $1\,\%$ in a fully ionized universe, even at the shortest quasar lifetimes, and grows with increasing IGM neutral fraction $\langle x_\mathrm{HI} \rangle$, peaking around $\langle x_\mathrm{HI} \rangle \sim 0.7$, and reaching values of up to $\sim7\,\%$ at the shortest lifetimes of $10^3 - 10^4\,\mathrm{yr}$. Notably, the scatter is not exclusively sourced by the distribution of ionized bubbles of the underlying reionization topology but also by the mere presence of density fluctuations which are in fact the \textit{only} source of scatter in an entirely neutral universe where $\langle x_\mathrm{HI} \rangle = 1$ and thus no topology variations are present.

In realistic settings where the intrinsic continuum of the quasar is unknown and has to be reconstructed too, the amount of scatter among different IGM transmission profiles seen in Figure~\ref{fig:transm_scatter_global_heatmap} is comparable to the $\sim 5\,\%$ uncertainties in the reconstructed quasar continuum \citep[c.f. Figure~6 in][]{hennawi2024}. As a result, this excess stochasticity in the IGM transmission model significantly adds to the total error budget on the reconstructed $\langle x_\mathrm{HI} \rangle$ and $t_\mathrm{Q}$ values \citep{kist2025}, whereas we could isolate it from the task of reconstructing the transmission profile if we were to work with a parameterization that more tightly captures the characteristic shape of IGM damping wings.

\subsection{Local parameterization}
\label{sec:local_scatter}

\begin{figure}
    \centering
	\includegraphics[width=\columnwidth]{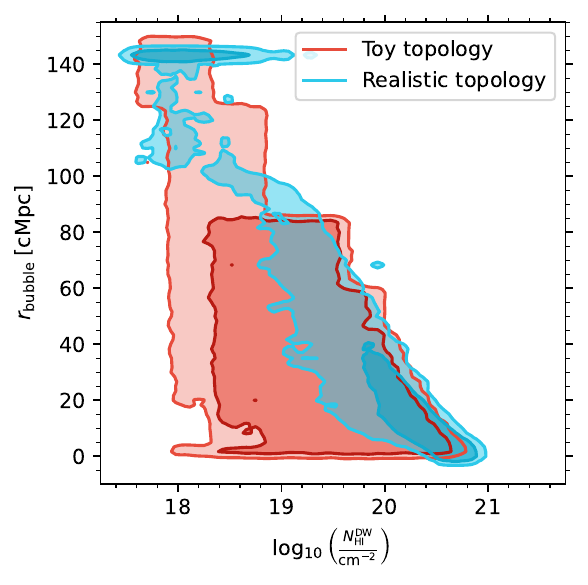}
    \caption{Distribution of our local summary statistics $N_\mathrm{HI}^\mathrm{DW}$ and $r_\mathrm{bubble}$ among all simulated IGM transmission profiles based on semi-numerical $x_\mathrm{HI}$ skewers (blue; see Section~\ref{sec:sims}), and analytical ones generated according to our toy bubble prescription (red; see Section~\ref{sec:bubble_model}). Dark shaded regions denote the $68\,\%$ contours, lighter ones the $95\,\%$ ones. Both distributions are smoothed via a kernel density estimation (KDE) with a Gaussian kernel of size $(0.05, 1.0)$ in $(\log_{10}N_\mathrm{HI}^\mathrm{DW}/\mathrm{cm}^{-2}, r_\mathrm{bubble}/\mathrm{cMpc})$ parameter space. The sharp edges of the toy model contours result from the fact that these profiles are simulated on a fixed grid of parameter values.}
    \label{fig:label_distr}

\end{figure}

\begin{figure*}
	\includegraphics[width=\textwidth]{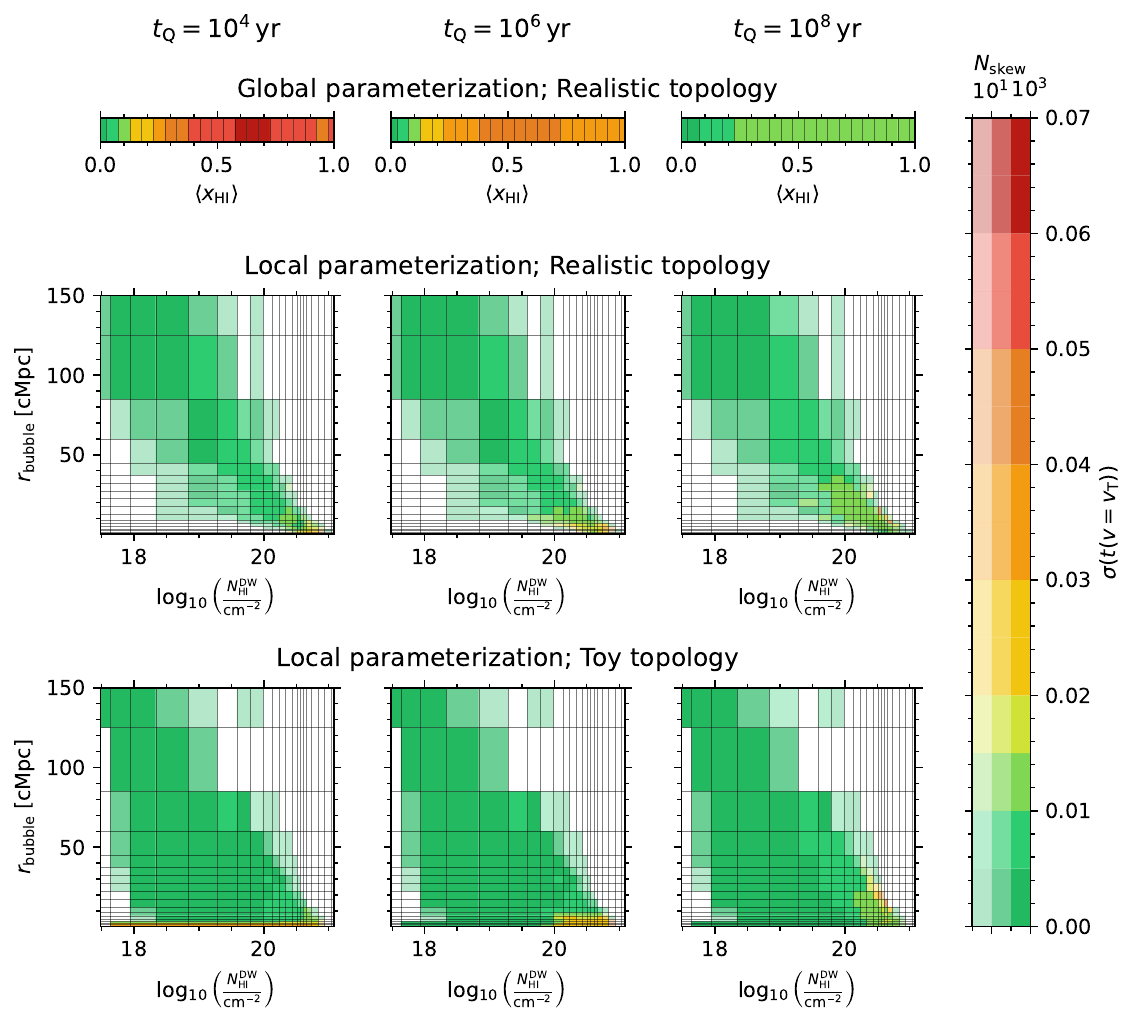}
    \caption{Sightline-to-sightline scatter of the IGM damping wing as a function of astrophysical parameter space for different parameterizations and topologies. The scatter is quantified as half the $68$-percentile width $\sigma(t(v = v_\mathrm{T}))$ of the distribution of IGM transmission values at $v = v_\mathrm{T}$ among all sightlines at a given location in parameter space, and at quasar lifetimes of $t_\mathrm{Q} = 10^4\,\mathrm{yr}$ (left column), $10^6\,\mathrm{yr}$ (middle column) and $10^8\,\mathrm{yr}$ (right column). \textit{Top row:} Profiles based on semi-numerical $x_\mathrm{HI}$ skewers (see Section~\ref{sec:sims}) in the global $\langle x_\mathrm{HI} \rangle$ parameterization. \textit{Middle row:} Same profiles as above, but in the local $(N_\mathrm{HI}^\mathrm{DW}, r_\mathrm{bubble})$ parameterization. \textit{Bottom row:} Profiles based on the analytical toy $x_\mathrm{HI}$ skewers (see Section~\ref{sec:bubble_model}) in the local $(N_\mathrm{HI}^\mathrm{DW}, r_\mathrm{bubble})$ parameterization. Bins which are occupied by a lower number of sightlines are more transparent, and white if not occupied at all.}
    \label{fig:transm_scatter_local_heatmap}
\end{figure*}

We now demonstrate that the scatter among individual IGM transmission profiles decreases significantly when labelling the profiles with the local summary statistics introduced in Section~\ref{sec:labels} in place of the global IGM neutral fraction $\langle x_\mathrm{HI} \rangle$.

To that end, we compute the distance-weighted HI column density $N_\mathrm{HI}^\mathrm{DW}$ as defined in Eq.~(\ref{eq:NHI_DW}) and the distance $r_\mathrm{bubble}$ to the first neutral bubble for each (pre-quasar) sightline and subsequently aggregate 
all post-quasar profiles according to these new labels. We fix the velocity offset and the integration range for $N_\mathrm{HI}^\mathrm{DW}$ as discussed in Sections~\ref{sec:NHI_DW} and \ref{sec:int_limits}. To compute $r_\mathrm{bubble}$, we smooth the $x_\mathrm{HI}$ field with a box-car filter of size $0.5\,\mathrm{cMpc}$ 
and determine $r_\mathrm{bubble}$ 
as the distance from the source where the smoothed $x_\mathrm{HI}$ field first becomes more than $50\,\%$ neutral.
This smoothing procedure makes sure that our $r_\mathrm{bubble}$ measurement identifies the first \textit{spatially extended} neutral bubble without being impacted by individual fluctuations in the $x_\mathrm{HI}$ field which do not have a notable impact on the damping wing shape. We find that we are not overly sensitive to the exact choices of these parameters.

The blue contours in Figure~\ref{fig:label_distr} depict the $68$ and $95\,\%$ regions of the distribution of $(N_\mathrm{HI}^\mathrm{DW}, r_\mathrm{bubble})$ values among all $852 \times 21$ sightlines.\footnote{Recall we are considering $852$ different sightlines at $21$ distinct global IGM neutral fraction values $\langle x_\mathrm{HI} \rangle$. Our $51$ different quasar lifetime values do not play a role as $N_\mathrm{HI}^\mathrm{DW}$ and $r_\mathrm{bubble}$ are defined on the pre-quasar topology.} As our $\langle x_\mathrm{HI} \rangle$ parameter grid uniformly covers the range between zero and one, this is the $(N_\mathrm{HI}^\mathrm{DW}, r_\mathrm{bubble})$ distribution that follows under the assumption that all neutral fraction values $0 \leq \langle x_\mathrm{HI} \rangle \leq 1$ are equally likely, i.e., in the absence of any knowledge about the reionization state of the IGM. We observe a clear degeneracy between $N_\mathrm{HI}^\mathrm{DW}$ and $r_\mathrm{bubble}$, where a higher column density $N_\mathrm{HI}^\mathrm{DW}$ implies a smaller $r_\mathrm{bubble}$ value and vice versa. This degeneracy 
arises due to the fact that a closer distance to the first neutral bubble clearly allows for more neutral material distributed within the range over which $N_\mathrm{HI}^\mathrm{DW}$ is computed. It is also apparent, however, that this is not a perfect degeneracy, 
and therefore the $r_\mathrm{bubble}$ statistic contains additional information not encoded in $N_\mathrm{HI}^\mathrm{DW}$.

To quantify this information, we define (unequally spaced) bins for the two labels and aggregate 
the transmission profiles accordingly. The binning is informed by the number of available skewers and the degree to which parameter variations impact the resulting IGM transmission profiles. Specifically, we choose a finer binning towards higher column densities and shorter distances to the first neutral bubble which cause more variation in the profiles due to increasingly strong damping wing absorption.

The scatter among the aggregated 
IGM transmission profiles in select parameter bins representative of the variations across the full parameter space is shown in Figure~\ref{fig:transm_scatter_NHI}. While a one-to-one comparison between individual bins in Figures~$\ref{fig:transm_scatter_xHI}$ and $\ref{fig:transm_scatter_NHI}$ is not possible due to the redefined bin labels, we did compute the effective line-of-sight averaged neutral fractions $\llangle x_\mathrm{HI} \rrangle_\mathrm{los}$ of the profiles in each parameter bin, and we depict in Figure~\ref{fig:transm_scatter_NHI} selected bins where the median $\llangle x_\mathrm{HI} \rrangle_\mathrm{los}$ matches the cosmic average $\langle x_\mathrm{HI} \rangle$ values shown in Figure~$\ref{fig:transm_scatter_xHI}$.\footnote{Recall that we denote cosmic averages with single angle brackets $\langle . \rangle$, and their $z=0$ values as $\langle . \rangle_{{}_0}$ throughout. We always denote line-of-sight averages with double angle brackets, where $\llangle . \rrangle_\mathrm{los}$ is the conventional line-of-sight average, and $\llangle . \rrangle_\mathrm{Lor}$ as defined in Eq.~(\ref{eq:Lorentzian_avg}) additionally involves a Lorentzian weighting function.}
With this in mind, we can certainly note that the scatter redward of the Lyman-$\alpha$ line in Figure~\ref{fig:transm_scatter_NHI} is consistently lower than in Figure~\ref{fig:transm_scatter_xHI}, implying that we identified a tighter parameterization of the characteristic shape of the IGM damping wing. Note that as expected, the scatter in the Lyman-$\alpha$ forest region is not impacted by the new parameterization.

For a more quantitative picture across the full range of parameter space, we summarize the $68$-percentile widths $\sigma(t(v=v_\mathrm{T}))$
of the distributions of IGM transmission values at $v=v_\mathrm{T}$ 
in all $(N_\mathrm{HI}^\mathrm{DW}, r_\mathrm{bubble})$ parameter bins in Figure~\ref{fig:transm_scatter_local_heatmap}.\footnote{Recall that we evaluate the IGM transmission scatter at $v=v_\mathrm{T}$, i.e., exactly at the reference velocity offset $v_\mathrm{T}$ used to determine $N_\mathrm{HI}^\mathrm{DW}$. Due to the strong pixel-by-pixel correlation across the entire IGM damping wing signature, this does \textit{not} mean we are underestimating the scatter, since this scatter can maximally contribute another $\sim0.5-1\,\%$ to $\sigma(t(v = v_\mathrm{T}))$ as demonstrated in Section~\ref{sec:one_param_family}.}
In each given row, we are showing from left to right the scatter at the three representative lifetimes of $t_\mathrm{Q} = 10^4$, $10^6$ and $10^8\,\mathrm{yr}$, respectively. For reference, the three upper panels depict the scatter in the global $\langle x_\mathrm{HI} \rangle$ parameterization,\footnote{Note that these are exactly the $t_\mathrm{Q} = 10^4$, $10^6$ and $10^8\,\mathrm{yr}$ slices of the heatmap in Figure~\ref{fig:transm_scatter_global_heatmap}.} whereas the three middle panels depict the scatter of the same profiles aggregated according to the $(N_\mathrm{HI}^\mathrm{DW}, r_\mathrm{bubble})$ parameterization. The three lower panels will be discussed in Section~\ref{sec:topology}. The transparency 
of the colorbar 
is proportional to the logarithm of
the number of profiles in each bin and thus represents the distribution of $(N_\mathrm{HI}^\mathrm{DW}, r_\mathrm{bubble})$ values which was already explicitly depicted in Figure~\ref{fig:label_distr}.

Remarkably, the transmission in the vast majority of bins in the local parameterization seen in the three middle panels of Figure~\ref{fig:transm_scatter_local_heatmap} varies by $\lesssim 1-2\,\%$, whereas it is consistently higher in the global parameterization depicted in the upper panels, reaching up to $\sim 7\,\%$ at $t_\mathrm{Q} = 10^4\,\mathrm{yr}$. This shows that our local summary statistics capture the characteristic shape of the IGM damping wing much more tightly than the global IGM neutral fraction $\langle x_\mathrm{HI} \rangle$.

At all lifetimes, however, we do observe a mild increase in scatter towards the highest column densities where the damping wing imprint gets more pronounced. This increased scatter is a result of the fact that we chose to define the two summaries on the \textit{pre}-quasar $x_\mathrm{HI}$ field. By fixing the integration limit for $N_\mathrm{HI}^\mathrm{DW}$ independently of quasar lifetime, we did not fully pay tribute to the variable radius of the ionization front $r_\mathrm{ion}$ where the integration effectively starts. Hence, by fixing $r_\mathrm{min} = 4\,\mathrm{cMpc}$, we are theoretically including too much or too little neutral material in the $N_\mathrm{HI}^\mathrm{DW}$ statistic, depending on the size of the ionized bubble the quasar has carved out, which in turn depends on the quasar lifetime,
as discussed in Section~\ref{sec:int_limits}. Nevertheless, the fact that the resulting scatter does not exceed $\sim 2\,\%$ even at the highest HI column densities shows that our approximation was a valid one, still leading to a tight parameterization of the \textit{post}-quasar IGM transmission field nearly independently of quasar lifetime.

\section{Towards a topology independent local IGM damping wing parameterization}
\label{sec:topology}

We introduced in the previous section a new three-parameter model for quasar IGM damping wings. We classified simulated IGM transmission profiles from a given reionization topology according to these labels and showed that the sightline-to-sightline scatter drastically decreases in this new parameterization 
as compared to the conventional \citep[e.g.][]{greig2017b, davies2018a, durovcikova2024, hennawi2024} one- or two-parameter model based on the global IGM neutral fraction $\langle x_\mathrm{HI} \rangle$ (and the quasar lifetime $t_\mathrm{Q})$. Specifically, by replacing $\langle x_\mathrm{HI} \rangle$ with the two \textit{local} summary statistics $N_\mathrm{HI}^\mathrm{DW}$ and $r_\mathrm{bubble}$, we were able to reduce the scatter in the damping wing region of the profile down to $\lesssim 1\,\%$ across the entire range of parameter space. In this section, we go beyond these results and demonstrate that our conclusions hold true largely independently of the underlying reionization topology.

This is because once our local summaries are set, the detailed distribution of neutral hydrogen along the line of sight, set by the reionization topology, does not matter anymore for the shape of the profile in the damping wing region of the spectrum. Instead, all relevant information about the damping wing shape is already contained in $N_\mathrm{HI}^\mathrm{DW}$ as Eq.~(\ref{eq:tau_DW_approx}) shows. The only remaining sources of scatter are the required approximations for this equality to hold, i.e., the finite integration limits of $N_\mathrm{HI}^\mathrm{DW}$, any differences between pre- and post-quasar density field that have not been mitigated by the choice of the lower integration limit and the inclusion of $r_\mathrm{bubble}$ as additional label, as well as any deviations of the damping wing shape from being a one-parameter family determined at a fixed reference distance (or velocity offset). However, all these contributions remain negligible as we showed in the previous section, neither do our conclusions change if the profiles originate from a different reionization topology as we will demonstrate here.

The scatter in the proximity zone region remains unaffected either because here the dominant source of sightline-to-sightline scatter is density fluctuations, 
while the $x_\mathrm{HI}$ field---only varying discontinuously between ionized and neutral patches with values close to zero and one---effectively acts as a window function for the fluctuations in the density field (see Eq.~(\ref{eq:nHI_gal}) and the example realization of these fields in Fig. \ref{fig:skewers}) which are already accounted for by considering a statistical sample of density skewers (independently of reionization topology). As a result, all effects of the bubble topology on the observed IGM transmission profile for a given sightline are already well captured in the two summary statistics $N_\mathrm{HI}^\mathrm{DW}$ and $r_\mathrm{bubble}$.

In Bayesian language, the mere effect of assuming a specific reionization topology is therefore imposing a corresponding prior on the parameters $N_\mathrm{HI}^\mathrm{DW}$ and $r_\mathrm{bubble}$ according to their distribution within that topology. For future damping wing analyses, this will allow one to perform a clear methodological distinction between 1) the topology-insensitive task of inferring the local damping wing statistics $(N_\mathrm{HI}^\mathrm{DW}, r_\mathrm{bubble}, t_\mathrm{Q})$, and 2) tying these constraints to a particular reionization topology, constraining not only the timing of reionization via the global IGM neutral fraction $\langle x_\mathrm{HI} \rangle$, but also its topology which remains unconstrained by $\langle x_\mathrm{HI} \rangle$ alone. It is immediately clear that measurements of $N_\mathrm{HI}^\mathrm{DW}$ (as a statistical moment of the $x_\mathrm{HI}$ field) and $r_\mathrm{bubble}$ for a statistical ensemble of objects carry additional topological information that we can extract and leverage to compare reionization models whose bubble topologies are different.

Here we demonstrate the topology independence of our labels\footnote{Again, the topology information is contained in the \textit{distribution} of these summaries while the summaries themselves can be defined and measured topology-independently.} in a simplified fashion by introducing a highly simplistic toy model to generate $x_\mathrm{HI}$ skewers analytically rather than extracting them from a realistic semi-numerical reionization topology. Specifically, we show that within our local parameter framework both models are statistically equivalent in terms of the resulting median IGM transmission profiles and scatter.

The specifics of the toy bubble model which we introduce in the following section are set keeping in mind the practical task of performing inference with respect to these new local summary statistics. Specifically, we introduced in \citet{hennawi2024} a fully Bayesian framework for inferring astrophysical model parameters from high-redshift quasar spectra. Although originally introduced to directly infer (among other parameters) the \textit{global} IGM neutral fraction, the framework is equally applicable to other parameters governing the IGM transmission field such as our local summary statistics. All which is needed is an estimate of the IGM transmission likelihood given the model parameters. Under the assumption of Gaussianity, this practically means we need to be able to estimate means and covariances of the IGM transmission field smoothly as a function of the model parameters $N_\mathrm{HI}^\mathrm{DW}$ and $r_\mathrm{bubble}$. 

As $N_\mathrm{HI}^\mathrm{DW}$ and $r_\mathrm{bubble}$ are derived quantities for IGM transmission profiles from the semi-numerical reionization simulation, the number of available sightlines can vary significantly from parameter bin to parameter bin. Especially for $(N_\mathrm{HI}^\mathrm{DW}, r_\mathrm{bubble})$ values that are rare in a given reionization topology, our covariance estimates can easily become so noisy that the likelihood function
is not guaranteed to vary smoothly as a function of $N_\mathrm{HI}^\mathrm{DW}$ and $r_\mathrm{bubble}$. While one could certainly beat down this noise with a massively increased number of simulations, building an analytical bubble model where the continuity 
of the IGM transmission field with respect to $N_\mathrm{HI}^\mathrm{DW}$ and $r_\mathrm{bubble}$ is intrinsically built in drastically reduces the number of required sightlines and facilitates the inference task for future applications, even though establishing the full inference pipeline would exceed the scope of this work. %

As far as this work is concerned, the bubble model provides us with a simple way of testing the sensitivity of our summary statistics to the underlying reionization topology by comparing the scatter of the IGM transmission profiles within these two very different topologies---the latter topology not even physically motivated. The fact that this scatter agrees remarkably well between the two demonstrates the topology-insensitivity of our labels and legitimizes the future use of the analytical bubble model for the purpose of astrophysical parameter inference.

\subsection{An analytic toy model for reionization bubbles}
\label{sec:bubble_model}

To generate IGM transmission profiles according to the toy bubble model which we will introduce now, we follow the same hybrid approach outlined in Section~\ref{sec:sims}, only replacing the semi-numerical $x_\mathrm{HI}$ skewers with analytically generated ones. This implies in particular that we use the same hydrodynamical skewers as described in that section and the same radiative transfer code to model the impact of the quasar ionizing radiation.

\subsubsection{Prescription of the toy bubble model}

We start by realizing that for a given density skewer, 
there exists a minimal and a maximal HI column density value $N_\mathrm{HI, min}^\mathrm{DW}$ and $N_\mathrm{HI, max}^\mathrm{DW}$ as computed according to Eq.~(\ref{eq:NHI_DW}). 
These values can only be achieved if the IGM is completely ionized or completely neutral (within the integration range), respectively. To determine $N_\mathrm{HI, min}^\mathrm{DW}$, we first have to compute the equilibrium ionization state for a fully ionized sightline based on the input UV background and temperature field, providing us with an $x_\mathrm{HI}$ field with values on the order of $O(10^{-3} - 10^{-4})$ which we can then use to compute $N_\mathrm{HI, min}^\mathrm{DW}$. To determine $N_\mathrm{HI, max}^\mathrm{DW}$, we subsequently set $x_\mathrm{HI} = 1$ at all distances greater than $r_\mathrm{bubble}$. Note that $N_\mathrm{HI, min}^\mathrm{DW}$ and $N_\mathrm{HI, max}^\mathrm{DW}$ differ from sightline to sightline, based on the specific realization of density fluctuations along the line of sight.

To achieve any column density value $\hat{N}_\mathrm{HI}^\mathrm{DW}$ between $N_\mathrm{HI, min}^\mathrm{DW}$ and $\leq N_\mathrm{HI, max}^\mathrm{DW}$, we start from the completely ionized sightline and add in neutral material until the desired $\hat{N}_\mathrm{HI}^\mathrm{DW}$ value is reached. 
We do so by subsequently adding in neutral bubbles $i$ of a minimum size $\Delta r_\mathrm{min}$ and a maximum size $\Delta r_\mathrm{max}$ originating at locations $r_i$ and growing them in the direction towards the observer.\footnote{We refer to $r_i$ as the 'origin' of this bubble and stress that it can only 'grow' in the direction away from the source.} Note that this procedure is by no means meant to resemble any physical processes occurring during reionization. It is a purely numerical procedure to ensure that
small changes in the model parameters $N_\mathrm{HI}^\mathrm{DW}$ and $r_\mathrm{bubble}$ result in small changes in the $x_\mathrm{HI}$ field and hence the IGM transmission profiles, allowing us to obtain smoothly varying means and covariances for future inference applications. To that end, we always follow the same sequence of locations $(r_i)_{i\in\mathbb{N}}$ when inserting new bubbles into a given sightline. In particular, we obtain the $x_\mathrm{HI}$ field for a given sightline and a desired parameter tuple $(\hat{N}_\mathrm{HI}^\mathrm{DW}, \hat{r}_\mathrm{bubble})$ according to the following procedure:
\begin{enumerate}
    \item Insert a neutral bubble of size $\Delta r_\mathrm{min}$ at the location $r_0 = \hat{r}_\mathrm{bubble}$, extending out to $\hat{r}_\mathrm{bubble} + \Delta r_\mathrm{min}$.
    \item Grow the current ($i$-th) bubble (originating at $r_i$) in the direction of the observer by subsequently setting
    $x_\mathrm{HI}$ pixel values adjacent to the outer edge of the bubble to unity. 
    In case a given pixel is already neutral because it overlaps with one of the previous neutral bubbles, this step has no effect.
    \item If the $i$-th bubble has reached a size of $\Delta r_\mathrm{max}$, insert a new bubble of size $\Delta r_\mathrm{min}$ originating at the next location $r_{i+1}$.
    \item If $r_{i+1} < \hat{r}_\mathrm{bubble}$, do not add the $(i+1)$-th bubble and instead skip to the $(i+2)$-th one because $\hat{r}_\mathrm{bubble}$ is by definition the distance to the neutral bubble \textit{closest} to the source. If necessary, repeat this step until a location is found where a bubble can be added.
    \item While $N_\mathrm{HI}^\mathrm{DW} < \hat{N}_\mathrm{HI}^\mathrm{DW}$, repeat steps (ii) - (iv).
\end{enumerate}

In this way, we are able to generate $x_\mathrm{HI}$ skewers on any desired $(N_\mathrm{HI}^\mathrm{DW}, r_\mathrm{bubble})$ grid, varying smoothly with both parameters. Note, however, that the $N_\mathrm{HI}^\mathrm{DW}$ ranges are physically restricted by $N_\mathrm{HI, min}^\mathrm{DW}$ and $N_\mathrm{HI, max}^\mathrm{DW}$ for each individual skewer, so it is not necessarily the case that there exists a realization of every sightline at each point in $(N_\mathrm{HI}^\mathrm{DW}, r_\mathrm{bubble})$ parameter space.

Moreover, for any given sightline, the maximal $N_\mathrm{HI}^\mathrm{DW}$ value changes as a function of $r_\mathrm{bubble}$. This is because a skewer with a certain neutral bubble distance $r_\mathrm{bubble}$ is by definition guaranteed to be ionized at $r < r_\mathrm{bubble}$, so the highest possible column density $N_\mathrm{HI, max}^\mathrm{DW}$ for a given sightline decreases with increasing $r_\mathrm{bubble}$. A subspace of the full $(N_\mathrm{HI}^\mathrm{DW}, r_\mathrm{bubble})$ parameter space is therefore \textit{physically inaccessible}, and this is determined \textit{exclusively} by the distribution 
of density fluctuations in the IGM, while entirely independent of any further assumptions about the reionization topology.

The prescription introduced in this section comes with two hyperparameters $\Delta r_\mathrm{min}$ and $\Delta r_\mathrm{max}$ determining the minimum and the maximum size of neutral bubbles that we insert into the sightlines. We choose these as $\Delta r_\mathrm{min} = 0.5\,\mathrm{cMpc}$ and $\Delta r_\mathrm{max} = 5.0\,\mathrm{cMpc}$, loosely informed by the sizes we empirically find for skewers in our semi-numerical reionization topology \citep[compare also][]{xu2017}, and noting that we are not sensitive to these exact choices.

\subsubsection{Generating IGM transmission profiles}

Using this prescription, we now generate a grid of IGM transmission profiles based on the same $852$ DLA-excluded density sightlines as in Section~\ref{sec:global_scatter}. We construct the grid in accordance with the $21$ and $18$ parameter bins used in Section~\ref{sec:local_scatter} to classify the realistic profiles with respect to $N_\mathrm{HI}^\mathrm{DW}$ and $r_\mathrm{bubble}$, respectively. Accordingly, we perform ionizing radiative transfer for the same lifetime values as in Section~\ref{sec:sims}, again focusing on the outputs after $t_\mathrm{Q} = 10^4$, $10^6$ and $10^8\,\mathrm{yr}$.

An example sightline can be found on the right-hand side of Figure~\ref{fig:skewers}. The resulting IGM transmission profile shown in the top panel is overlaid 
in red with the one we obtained using a semi-numerical $x_\mathrm{HI}$ skewer (blue; left-hand panels) as considered in Section~\ref{sec:sims}.
For optimal comparability to the realistic reionization topology, we used
the same underlying density skewer 
and matched the parameter values, i.e., both the semi-numerical sightline (blue) and the analytical toy-sightline (red) have $\log_{10}N_\mathrm{HI}^\mathrm{DW}/\mathrm{cm}^{-2} = 20.1$, $r_\mathrm{bubble} = 8\,\mathrm{cMpc}$, and $t_\mathrm{Q} = 10^6\,\mathrm{yr}$, where the semi-numerical sightline originates from a topology with a global IGM neutral fraction of $\langle x_\mathrm{HI} \rangle = 0.65$. The pre-quasar versions of all physical fields are shown in dark, and those of the IGM transmission field in lighter colors. %

Despite the locally very different $x_\mathrm{HI}$ fields, the resulting IGM transmission profiles look almost identical in the smooth damping wing region redward of the Lyman-$\alpha$ line, and even their proximity zones agree relatively well due to the matching density profiles giving rise to similar Lyman-$\alpha$ absorption signatures. The actual location of the quasar ionization front, however, differs notably among the two examples due to the larger amounts of neutral gas near the quasar for the semi-numerical sightline. Its proximity zone therefore does not extend quite as far out as that of the toy sightline. However, as we will show statistically in the next section, such differences in the proximity zone region due to the distribution of neutral material are entirely outweighed by the differences caused by density fluctuations alone. The damping wing region remains completely unaffected, as our $N_\mathrm{HI}^\mathrm{DW}$ label by construction accounts for neutral material along the \textit{entire} line of sight with the correct weighting. This suggests that our parameterization well captures the characteristic shape of the IGM damping wing, even in a case where the sightlines originate from highly different reionization topologies. We proceed in the following section with a statistical confirmation of these sightline-based observations.

\subsection{Comparison to a realistic reionization topology}

\begin{figure*}
	\includegraphics[width=0.95\textwidth]{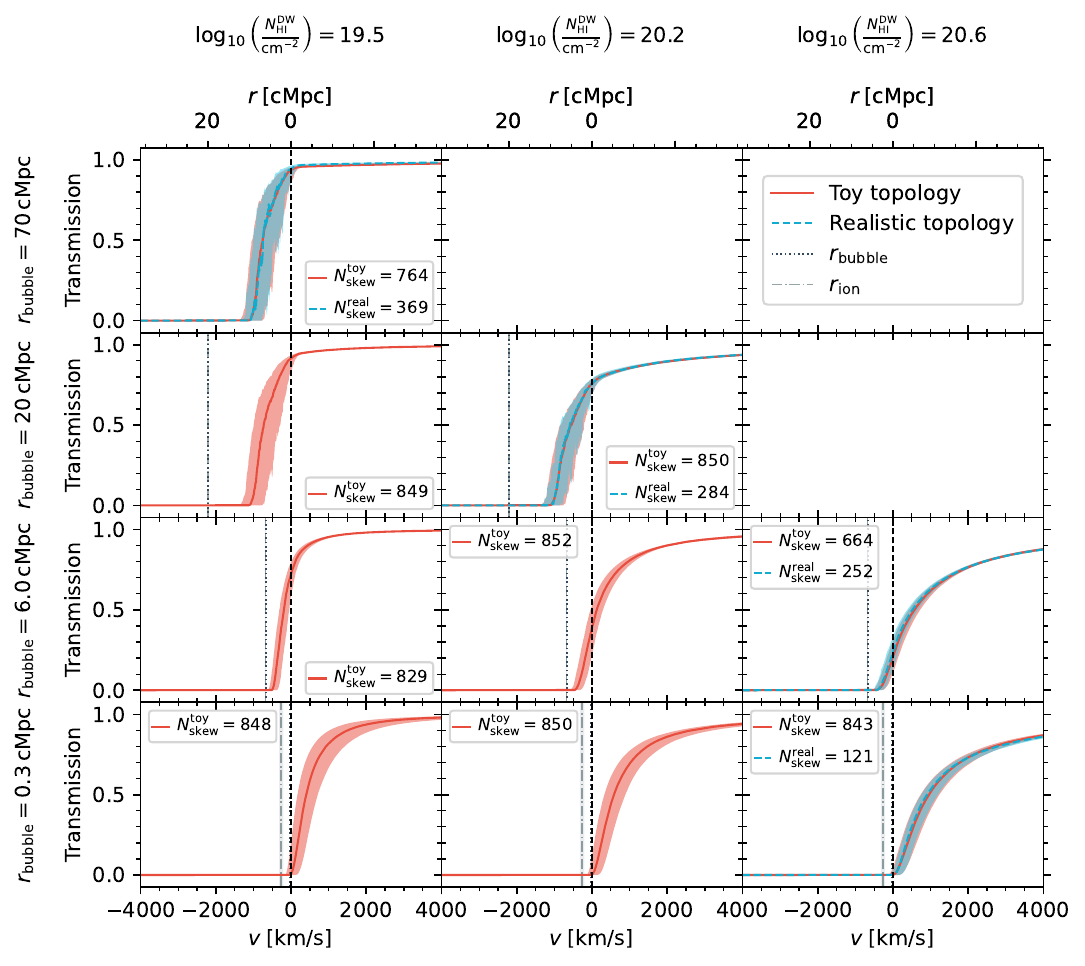}
    \caption{Median and $68$-percentile scatter of simulated IGM transmission profiles in the local $(N_\mathrm{HI}^\mathrm{DW}, r_\mathrm{bubble})$ parameterization at a quasar lifetime of $t_\mathrm{Q} = 10^4\,\mathrm{yr}$, comparing profiles based on semi-numerical $x_\mathrm{HI}$ skewers (blue; see Section~\ref{sec:sims}), and analytical ones generated according to our toy bubble prescription (red; see Section~\ref{sec:bubble_model}). The profiles are shown on a representative grid of $(N_\mathrm{HI}^\mathrm{DW}, r_\mathrm{bubble})$ parameter values. For clarity, bins which are occupied by no more than $30$ sightlines are omitted.}
    \label{fig:toy_transm_scatter_logtQ4}
\end{figure*}

\begin{figure*}
	\includegraphics[width=0.95\textwidth]{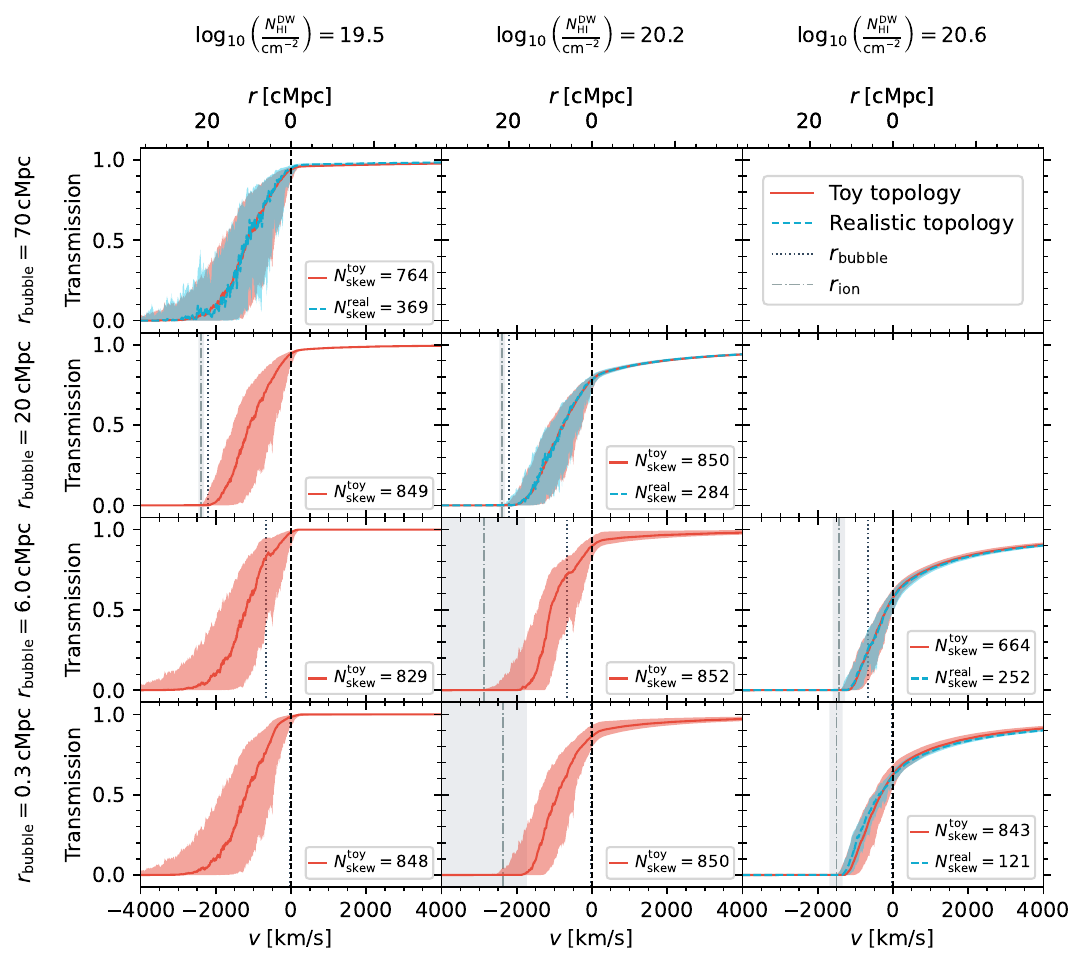}
    \caption{Like Figure~\ref{fig:toy_transm_scatter_logtQ4}, but for a quasar lifetime of $t_\mathrm{Q} = 10^6\,\mathrm{yr}$.}
    \label{fig:toy_transm_scatter_logtQ6}
\end{figure*}

\begin{figure*}
	\includegraphics[width=0.95\textwidth]{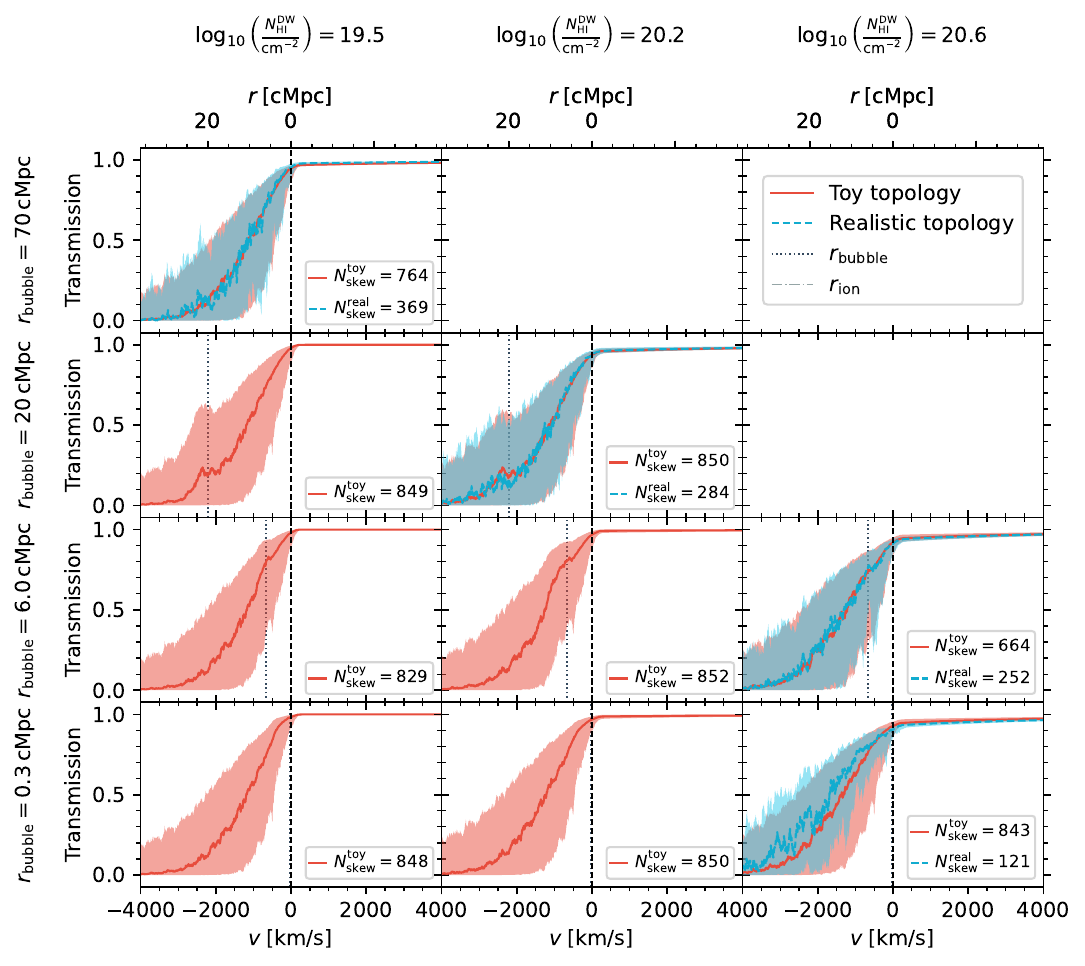}
    \caption{Like Figure~\ref{fig:toy_transm_scatter_logtQ4}, but for a quasar lifetime of $t_\mathrm{Q} = 10^8\,\mathrm{yr}$.}
    \label{fig:toy_transm_scatter_logtQ8}
\end{figure*}

We now investigate the scatter in the IGM transmission profiles based on our newly introduced toy prescription, and compare it to that found in Section~\ref{sec:local_scatter} in the context of a more realistic semi-numerical reionization topology. We depict the median and the $68$-percentile scatter of the IGM transmission profiles at representative locations in $(N_\mathrm{HI}^\mathrm{DW}, r_\mathrm{bubble})$ parameter space in Figures~\ref{fig:toy_transm_scatter_logtQ4}, \ref{fig:toy_transm_scatter_logtQ6} and \ref{fig:toy_transm_scatter_logtQ8}. The number of sightlines based on which the percentiles are computed are denoted in each panel. The three different figures show the same sets of profiles at quasar lifetimes of $t_\mathrm{Q} = 10^4$, $10^6$ and $10^8\,\mathrm{yr}$, respectively. In addition to the profiles based on our toy topology (red), we also show the
ones from the realistic reionization topology (blue; see also Figure~\ref{fig:transm_scatter_NHI}), aggregated into the same parameter bins as introduced in Section~\ref{sec:local_scatter}.\footnote{Note that all toy profiles are simulated at a unique set of parameter values for each given bin, whereas the semi-numerical ones are aggregated from the entire parameter range covered by this bin.} For clarity, we omit profiles in noise-dominated parameter bins with no more than $30$ sightlines.

\subsubsection{Physically accessible regions of parameter space}

The first thing of note is that our toy profiles extends over a significantly larger region in $(N_\mathrm{HI}^\mathrm{DW}, r_\mathrm{bubble})$ parameter space as compared to the realistic ones, as we can immediately see from the contours depicted in Figure~\ref{fig:label_distr}. Both these models show a non-trivial dependence on the parameter values due to the fact that not all regions of parameter space are accessible given the distribution of density fluctuations in the IGM. However, the toy bubble model explores significantly larger regions in parameter space than the more realistic semi-numerical model. This shows that there are parameter configurations which are theoretically possible for a given sightline but which do not necessarily appear in our simulated reionization topology.

This mainly concerns combinations of small HI column densities $N_\mathrm{HI}^\mathrm{DW}$ with short neutral bubble distances $r_\mathrm{bubble}$. Such configurations are increasingly hard to achieve the smaller $r_\mathrm{bubble}$, as neutral material closer to the source gets upweighted according to the Lorentzian weighting function. A very nearby neutral bubble therefore has to be extremely small in extent as to not exceed a given HI column density $N_\mathrm{HI}^\mathrm{DW}$---a configuration that hardly occurs in the realistic reionization topology. The relatively tight correlation between $N_\mathrm{HI}^\mathrm{DW}$ and $r_\mathrm{bubble}$ seen for the blue contours in Figure~\ref{fig:label_distr} is therefore specific to this topology, and not necessarily a generic feature of other ones.
In any case, as seen in the three lower panels of Figure~\ref{fig:transm_scatter_local_heatmap}, the IGM transmission scatter as quantified by the $68$-percentile widths $\sigma(t(v=v_\mathrm{T}))$ remains similarly low even in these regions of parameter space which remain unexplored by the realistic topology. Our parameterization therefore tightly captures the shape of the IGM damping wing in \textit{all} regions of parameter space which are physically allowed.

The unpopulated regions of parameter space at high column densities with large neutral bubble distances, on the other hand, are a hard physical constraint, independent of reionization topology. As argued in Section~\ref{sec:bubble_model}, the maximum column density $N_\mathrm{HI}^\mathrm{DW}$ for a given sightline at a given neutral bubble distance $r_\mathrm{bubble}$ can only be achieved if the skewer is fully neutral starting at $r\geq r_\mathrm{bubble}$ (and, by definition, fully ionized at $r < r_\mathrm{bubble}$). This maximum value therefore decreases with increasing $r_\mathrm{bubble}$ for a given sightline. Parameter combinations $(N_\mathrm{HI}^\mathrm{DW}, r_\mathrm{bubble})$ where we cannot find \textit{any} sightline that reaches the corresponding column density are therefore \textit{physically excluded} since even the strongest density fluctuations do not suffice anymore to achieve this HI column density $N_\mathrm{HI}^\mathrm{DW}$. Again, this statement is solely based on the underlying distribution of density fluctuations and independent of the topology of reionization.

\subsubsection{Agreement between realistic and toy bubble model}

Moving on to the comparison of the IGM transmission profiles between the realistic and the toy topology, we observe a remarkable agreement throughout nearly all bins that are populated by both models, across the entire lifetime range, as can be seen in the panels of Figures~\ref{fig:toy_transm_scatter_logtQ4}, \ref{fig:toy_transm_scatter_logtQ6} and \ref{fig:toy_transm_scatter_logtQ8} where profiles from both models are available. The only bins where differences become visible are the bottom right ones in each figure where $\log_{10} N_\mathrm{HI}^\mathrm{DW}/\mathrm{cm}^{-2} = 20.6$ and $r_\mathrm{bubble} = 0.3\,\mathrm{cMpc}$.\footnote{Due to the lack of a star formation prescription, the Nyx simulations do not correctly capture the circumgalactic medium of massive halos, and therefore we exclude the first $0.3\,\mathrm{cMpc}$ of all sightlines. Hence $r_\mathrm{bubble} = 0.3\,\mathrm{cMpc}$ is the shortest neutral bubble distance possible in our simulations. Note that this does not imply our simulated transmission profiles are unphysical, as material this close to the quasar is certainly guaranteed to be ionized away, even for the shortest quasar lifetimes.} While both models still perfectly agree at the lowest lifetime of $t_\mathrm{Q} = 10^4\,\mathrm{yr}$ (see Figure~\ref{fig:toy_transm_scatter_logtQ4}), the profiles increasingly differ at longer lifetimes of $t_\mathrm{Q} = 10^6$ and $10^8\,\mathrm{yr}$ as seen in Figures~\ref{fig:toy_transm_scatter_logtQ6} and \ref{fig:toy_transm_scatter_logtQ8}.

More specifically, we see that while the profiles still agree reasonably well in the damping wing region of the spectrum, the proximity zone transmission of the realistic profiles is systematically higher than that of the toy ones. The reason for this is related to the fact that in the realistic semi-numerical reionization topology, $r_\mathrm{bubble}$ values of $0.3\,\mathrm{cMpc}$ are in practice only found for completely neutral sightlines, since a small $r_\mathrm{bubble}$ value always implies a large column density $N_\mathrm{HI}^\mathrm{DW}$ as can be see in Figure~\ref{fig:label_distr}. On the other hand, our toy model also allows for many sightlines with a first neutral bubble at $r_\mathrm{bubble} = 0.3\,\mathrm{cMpc}$ and a number of ionized patches farther away from the source, leading to lower column densities, and hence the contours extending out to much smaller $N_\mathrm{HI}^\mathrm{DW}$. These configurations are extremely rare in realistic reionization topologies as the regions around the most massive halos where our quasars are placed are among the first ones to be reionized, whereas no such constraint applies to our toy bubble model where we only place the first neutral bubble at the desired location and subsequently add more neutral bubbles at \textit{any} location behind it until a desired column density is reached.

Having available this broader range of $x_\mathrm{HI}$ 
skewers at $r_\mathrm{bubble} = 0.3\,\mathrm{cMpc}$ means that we can achieve a given HI column density 
$N_\mathrm{HI}^\mathrm{DW}$ for many additional density sightlines. 
These additional sightlines are generically not as underdense as the ones that achieve a given $N_\mathrm{HI}^\mathrm{DW}$ value in combination with a completely neutral sightline. This is because \textit{both} $x_\mathrm{HI}$ and $\Delta$ contribute to $N_\mathrm{HI}^\mathrm{DW}$ (see Eq.~(\ref{eq:NHI_DW})), and if $x_\mathrm{HI}$ is equal to unity, $\Delta$ must be comparatively low. Vice versa, the additional, globally more ionized, skewers from our toy topology can be combined with sightlines less underdense than those required for completely neutral skewers to achieve the same column density $N_\mathrm{HI}^\mathrm{DW}$.
The more typical densities 
present in these additional sightlines in return lead to less pronounced transmission spikes in the proximity zone region which become increasingly apparent the larger the proximity zone sizes, i.e., the longer the quasar lifetimes, as the progression of the bottom right panels of Figures~\ref{fig:toy_transm_scatter_logtQ4}, \ref{fig:toy_transm_scatter_logtQ6} and \ref{fig:toy_transm_scatter_logtQ8} shows. The damping wing region on the other hand remains largely unaffected; if any, we observe a slight increase in damping wing absorption strength for the realistic sightlines as they are globally the most neutral ones. 
However, our $N_\mathrm{HI}^\mathrm{DW}$ statistic \textit{by construction} already accounts for the competing effects between $x_\mathrm{HI}$ and $\Delta$ with the correct distance weighting, and hence the damping wing strength still agrees well among both models. The remaining small differences among the two models can be attributed to differences in the neutral hydrogen content in front of our fixed lower integration limit of $r_\mathrm{min} = 4\,\mathrm{cMpc}$.

While Figures~\ref{fig:toy_transm_scatter_logtQ4}, \ref{fig:toy_transm_scatter_logtQ6} and \ref{fig:toy_transm_scatter_logtQ8} only show the scatter of the IGM transmission profiles in select parameter bins, we verified that the effect described above only becomes relevant in parameter bins with $r_\mathrm{bubble} = 0.3\,\mathrm{cMpc}$, while both models perfectly agree across the entire rest of $(N_\mathrm{HI}^\mathrm{DW}, r_\mathrm{bubble})$ parameter space (modulo a somewhat increased statistical noise level for the realistic topology in bins which are more scarcely populated). The prior volume where our toy prescription introduces a small bias therefore remains minimal, and this bias certainly remains subdominant to other major sources of uncertainty in realistic inference settings where e.g. the intrinsic continuum of the quasar has to be reconstructed as well \citep{kist2025}. Correspondingly, the $68$-percentile widths $\sigma(t(v=v_\mathrm{T}))$ of the IGM transmission distributions coincide remarkably well between the two topologies in all regions of parameter space, as can be seen by comparing the middle to the lower panels of Figure~\ref{fig:transm_scatter_local_heatmap}.

Overall, this leads us to the striking conclusion that the statistical properties (median and $68$-percentile scatter) of the IGM transmission profiles are largely insensitive to the peculiarities of \textit{any} given reionization topology. Again, we compared transmission profiles from two entirely different reionization bubble models---the latter of which is not even physically motivated---and yet, no differences are recognizable across the largest parts of parameter space. Even the remaining differences at $r_\mathrm{bubble} = 0.3\,\mathrm{cMpc}$ are a mere relic of the unphysical fact that our toy bubble model does not account for regions around the most massive halos reionizing early. In future applications, we could alleviate this effect by modifying our toy model to respect other spatial correlations between $x_\mathrm{HI}$ and $\Delta$ such as their average radial profile in addition to $N_\mathrm{HI}^\mathrm{DW}$ and $r_\mathrm{bubble}$ instead of randomly laying down the neutral patches. Our statement about topology independence of our new set summary statistics therefore remains unconditionally true when comparing profiles from any two realistic reionization topologies where this property is accounted for.

\subsubsection{Parameter dependence of the IGM transmission profiles at short quasar lifetimes}

After these general considerations, we now investigate how the shape of the IGM transmission profiles changes as a function of $N_\mathrm{HI}^\mathrm{DW}$ and $r_\mathrm{bubble}$. We start by considering the short lifetime ($t_\mathrm{Q} = 10^4\,\mathrm{yr}$) profiles depicted in Figure~\ref{fig:toy_transm_scatter_logtQ4}. As expected, higher HI column densities generally cause stronger IGM damping wing imprints. Also our second summary statistic $r_\mathrm{bubble}$ has a clear impact on the shape of the IGM transmission profiles, even though it is defined as the distance to the first neutral bubble in the \textit{pre}-quasar topology.
The first obvious reason for this is the fact that $N_\mathrm{HI}^\mathrm{DW}$ is defined to be insensitive to any kind of structure in front of our lower integration limit $r_\mathrm{min} = 4\,\mathrm{cMpc}$. For quasars whose first neutral bubble is located at $r_\mathrm{bubble} < r_\mathrm{min}$, and whose lifetime is not long enough to ionize away this neutral bubble, $r_\mathrm{bubble}$ is the only label that can carry information about the remaining nearby neutral material which $N_\mathrm{HI}^\mathrm{DW}$ by definition cannot capture.

However, we see in Figure~\ref{fig:toy_transm_scatter_logtQ4} that the IGM transmission profiles still carry information about $r_\mathrm{bubble}$ even if $r_\mathrm{bubble} > r_\mathrm{min}$. %
This is because on average, a short-lived quasar which has only been shining for a few thousands of years has not yet had sufficient time to carve out an ionized region extending (significantly) beyond the start of the first neutral bubble (regardless how its original location $r_\mathrm{bubble}$ relates to $r_\mathrm{min}$), and so $r_\mathrm{bubble}$ constitutes a hard upper limit for the location of its ionization front $r_\mathrm{ion}$ (marked with a vertical dash-dotted line in each panel).\footnote{The location of the ionization front $r_\mathrm{ion}$ is measured empirically based on all toy sightlines which are available in a given parameter bin. To that end, we take the same algorithm used to determine $r_\mathrm{bubble}$ for the pre-quasar topology, and apply it to the post-quasar topology instead. The dash-dotted line marks the median $r_\mathrm{ion}$ of all toy profiles in a given panel, and the grey-shaded region the $68$-percentile scatter.} The transmission therefore certainly drops to zero at or prior to $r_\mathrm{bubble}$ (dotted vertical lines), and hence $r_\mathrm{bubble}$ carries additional information about the shape of the profile in this region, clearly supplementing the information about the red-side damping wing shape which is primarily encoded in the $N_\mathrm{HI}^\mathrm{DW}$ statistic. This can be seen by comparing the panels within a given row (i.e., at fixed $N_\mathrm{HI}^\mathrm{DW}$) in Figure~\ref{fig:toy_transm_scatter_logtQ4}.
In these regions of parameter space where the transmission drops to zero close to $r_\mathrm{bubble}$, this label therefore encodes similar information as the velocity realignment point proposed in \citet{chen2024} and \citet{keating2024a} (when considered at a fixed quasar lifetime $t_\mathrm{Q}$). %

On the other hand, no differences are recognizable between the profiles in the two upper rows of Figure~\ref{fig:toy_transm_scatter_logtQ4} where all transmission is extinguished at a parameter-independent location significantly closer to the source than the position $r_\mathrm{bubble}$ of the first neutral bubble. 
As the IGM is fully ionized up to the latter point, the Gunn-Peterson optical depth $\tau_\mathrm{GP}$ in this region is inversely proportional to the photoionization rate $\Gamma_\mathrm{QSO}$ of the quasar, i.e., $\tau_\mathrm{GP} \sim 1/\Gamma_\mathrm{QSO}$. Since the quasar's photoionization rate decreases as the inverse square of the distance $r$ from the quasar, $\Gamma_\mathrm{QSO} \sim 1/r^2$, all transmission eventually gets suppressed beyond a certain distance, even if the quasar ionization front has already passed through.
As a result, profiles with $r_\mathrm{bubble}$ values exceeding this distance are entirely degenerate. This degeneracy would be ameliorated if we were to consider a brighter quasar whose $\Gamma_\mathrm{QSO}$ would be higher and hence this distance would be located farther outwards from the source.

\subsubsection{Parameter dependence of the IGM transmission profiles at intermediate to long quasar lifetimes}

The sensitivity to $r_\mathrm{bubble}$ decreases as we move to older objects with lifetimes of $t_\mathrm{Q} = 10^6$ or $10^8\,\mathrm{yr}$ such as depicted in Figures~\ref{fig:toy_transm_scatter_logtQ6} and \ref{fig:toy_transm_scatter_logtQ8}. 
Such quasars have had a sufficient amount of time to carve out a large ionized region around themselves, 
in most cases exceeding the first pre-quasar neutral bubble location and hence, the transmission does not sharply drop to zero at $r_\mathrm{bubble}$. 
Instead, $r_\mathrm{bubble}$ is located somewhere within the proximity zone of most such quasars, and as such is hardly reconstructable from a given IGM transmission profile. 
Note that in the two panels of Figure~\ref{fig:toy_transm_scatter_logtQ6} where $r_\mathrm{bubble} = 20\,\mathrm{cMpc}$
(and $t_\mathrm{Q} = 10^6\,\mathrm{yr}$), we still do observe a similar drop-off close to $r_\mathrm{bubble}$ as we did for shorter neutral bubble distances at $t_\mathrm{Q} = 10^4\,\mathrm{yr}$. This shows that at a lifetime of $t_\mathrm{Q} = 10^6\,\mathrm{yr}$, a neutral bubble located at $r_\mathrm{bubble} = 20\,\mathrm{cMpc}$ still constitutes a comparably hard limit for the quasar ionization front which comes to a halt shortly thereafter, whereas closer neutral bubbles easily get ionized away such that $r_\mathrm{ion}$ significantly exceeds $r_\mathrm{bubble}$.

Remarkably, however, even in the case where the first pre-quasar neutral bubble is ionized away, we statistically still do observe a remnant signature of $r_\mathrm{bubble}$ in form of an excess transmission bump starting at exactly this location. This feature can most clearly be identified in the central panels of Figures~\ref{fig:toy_transm_scatter_logtQ6} and \ref{fig:toy_transm_scatter_logtQ8} where both median and scatter are elevated starting precisely at $r_\mathrm{bubble}$. Most notably, a noisier version of this bump can even be identified for the semi-numerical IGM transmission profiles in the $(\log_{10}N_\mathrm{HI}^\mathrm{DW}/\mathrm{cm}^{-2}, r_\mathrm{bubble}/\mathrm{cMpc}) = (20.2, 20)$ panel of Figure~\ref{fig:toy_transm_scatter_logtQ8}. We attribute this feature to photoelectric heating of the (neutral) IGM by the hard quasar spectrum.

The mechanism can be understood by returning to the example sightline depicted in Figure~\ref{fig:skewers}: comparing the temperature fields before and after the quasar has turned on, we see that the initially neutral regions get heated significantly (bottom row) after getting ionized away by the quasar. This is because at large optical depth, the soft photons all get absorbed, and the optical depth of the harder photons is also high so they get absorbed as well, causing photoelectric heating up to $40\,000\,\mathrm{K}$ (c.f. also Figure~6 of \citet{davies2023b}). As a result, the transmission is enhanced in these regions. Since this significant amount of heating is restricted to initially neutral patches which recently got ionized by the quasar, the \textit{pre}-quasar $x_\mathrm{HI}$ topology is still encoded in the \textit{post}-quasar temperature field, and hence also leaves an imprint on the IGM transmission field. For an individual sightline, however, this signature is hard to disentangle from the stochasticity of the Lyman-$\alpha$ absorption features in the proximity zone of the quasar due to density fluctuations, even more so when the continuum is unknown, and in the presence of observational noise. Yet, our $(N_\mathrm{HI}^\mathrm{DW}, r_\mathrm{bubble})$ parameterization allows us to statistically account for this feature when performing astrophysical parameter inference. We will explore this possibility in future work.

\section{Conclusions}
\label{sec:conclusions}

We introduced in this work a novel three-parameter model that tightly captures the characteristic shape of quasar IGM damping wings. As an alternative to the common parameterization based on the global IGM neutral fraction $\langle x_\mathrm{HI} \rangle$, we defined two new \textit{local} summary statistics quantifying the neutral hydrogen content along the sightline from the quasar \textit{before} it started shining: 1) the distance-weighted HI column density $N_\mathrm{HI}^\mathrm{DW}$, and 2) the distance $r_\mathrm{bubble}$ between the quasar and the first neutral patch. We supplemented these two local measures of the \textit{pre}-quasar neutral topology with the quasar lifetime $t_\mathrm{Q}$ as a third parameter encapsulating the effects of the ionizing quasar radiation.

Since quasars are typically going to ionize away all surrounding neutral gas within the first few $\mathrm{cMpc}$, we found that
the damping wing is most sensitive to the distance-weighted HI column density $N_\mathrm{HI}^\mathrm{DW}$ starting at $4\,\mathrm{cMpc}$ from the source, and integrated over a range of $100\,\mathrm{cMpc}$. By adding a Lorentzian weighting function (associated to a fixed reference distance $r_\mathrm{T} = 18\,\mathrm{cMpc}$) to the column density integral, we account for the effect of the Lyman-$\alpha$ cross section $\sigma_\alpha$ in the optical depth integral, assuring a maximally tight parameterization of the IGM damping wing at the spectral pixel corresponding to $r_\mathrm{T}$.
By demonstrating that the damping wing essentially constitutes a one-parameter family, we showed that $N_\mathrm{HI}^\mathrm{DW}$ tightly parameterizes the damping wing imprint not only at this specific location but across the \textit{entire} spectral range.

We introduced as a second summary statistic the distance $r_\mathrm{bubble}$ from the source to the first neutral patch in the \textit{pre}-quasar topology, and showed that it still encodes information about \textit{post}-quasar IGM transmission profiles, and that this information is complementary to $N_\mathrm{HI}^\mathrm{DW}$. We related the definitions of our summary statistics back to the labels recently proposed in \citet{chen2024}, \citet{keating2024a}, and \citet{mason2025}, and argued that by starting the $N_\mathrm{HI}^\mathrm{DW}$ integration directly at the location of the source, our parameterization is also applicable in the context of IGM damping wings towards galaxies.

We simulated realistic IGM transmission profiles and compared their scatter in the damping wing region of the spectrum at fixed parameter values within both the global and the local parameterization. We found that due to the stochastic distribution of neutral patches during reionization and density fluctuations in the IGM, the $68$-percentile scatter of the IGM transmission values at $v_\mathrm{T} = 2000\,\mathrm{km}/\mathrm{s}$ in the global parameterization can be as large as $7\,\%$ (at a quasar lifetime of $t_\mathrm{Q} = 10^4\,\mathrm{yr}$), and we demonstrated that this scatter decreases down to $\lesssim 1\,\%$ across the entire range of physical parameter space when aggregating the same sightlines according to our local parameterization.

We introduced a simple numerical prescription to generate synthetic HI density profiles at any desired location in $(N_\mathrm{HI}^\mathrm{DW}, r_\mathrm{bubble})$ parameter space, varying smoothly as a function of both these parameters. Even though the procedure does not attempt to describe any physical processes, it allowed us to demonstrate the robustness of our parameterization against the choice of reionization model since it results in a significantly different bubble topology.
We observed an exceptional agreement between the realistic and the toy profiles at almost every location in physical parameter space. Small differences between the models only became apparent for profiles with a neutral bubble directly next to the quasar, which we could trace back to the fact that our toy prescription does not pay tribute to the fact that reionization takes place inside out. As this would be accounted for by any realistic reionization model, the remarkable overall agreement between the two models demonstrates the topology-insensitivity of our parameterization.

In addition, this agreement legitimizes the future use of our toy prescription for the purpose of astrophysical parameter inference, decoupling 1) the topology-insensitive task of inferring the local damping wing statistics introduced in this work, and 2) tying these constraints to a specific reionization model, resulting in near-optimal constraints not only on the global timing of reionization, but also the reionization topology, hitherto unconstrained with quasar IGM damping wings. In this context, all assumptions about the reionization model can be encoded in a prior on $(N_\mathrm{HI}^\mathrm{DW}, r_\mathrm{bubble})$, determined by the distribution of these parameters within the model of interest. This clear separation will facilitate future inference endeavors, and even the comparison of different reionization models, harnessing \textit{all} the information encapsulated in quasar IGM damping wings.

\section*{Acknowledgements}

We acknowledge helpful conversations with the ENIGMA group at UC Santa Barbara and Leiden University and would especially like to thank Shane Bechtel and Benjamin Snyder for comments on an early version of this manuscript.
This work made use of \texttt{NumPy} \citep{harris2020},  \texttt{SciPy} \citep{virtanen2020}, \texttt{Astropy} \citep{astropy_collaboration2013, astropy_collaboration2018, astropy_collaboration2022}, \texttt{h5py} \citep{collette2013}, \texttt{Matplotlib} \citep{hunter2007}, and \texttt{IPython} \citep{perez2007}.
TK and JFH acknowledge support from the European Research Council (ERC) under the European Union’s Horizon 2020 research and innovation program (grant agreement No 885301). JFH acknowledges support from NSF grant No. 2307180.

\section*{Data Availability}

The derived data generated in this research will be shared on reasonable requests to the corresponding author.

\bibliographystyle{mnras}
\bibliography{main} %

\appendix

\section{On the relation between $\tau_\mathrm{DW}$ and $N_\mathrm{HI}^\mathrm{DW}$}
\label{app:tau_vs_NHI}
We introduced in Section~\ref{sec:labels} the distance-weighted HI column density $N_\mathrm{HI}^\mathrm{DW}$ as a summary statistic of the local pre-quasar neutral hydrogen content along the line of sight from the source.
By comparing $N_\mathrm{HI}^\mathrm{DW}$ as defined in Eq.~(\ref{eq:NHI_DW}) to the pre-quasar damping wing optical depth
\begin{equation}
\label{eq:tau_DW_pre}
    \tau_\mathrm{DW}^\mathrm{pre}(\lambda_\mathrm{rest}) = \int_{0}^{R(z_\mathrm{QSO})} n_\mathrm{HI}^\mathrm{pre}(R) \cdot \sigma_\alpha\left(\frac{1+z_\mathrm{QSO}}{1+z(R)}\,\lambda_\mathrm{rest}\right) \; \mathrm{d}R,
\end{equation}
we motivate in this section why this summary statistic is a near-optimal means to reduce the IGM transmission scatter redward of the Lyman-$\alpha$ line. Specifically, we show that in the limit where the distance-weighted average $\llangle x_\mathrm{HI} \cdot \Delta \rrangle_{\mathrm{Lor}}$ as defined in Eq.~(\ref{eq:Lorentzian_avg}) is computed along the \textit{entire} line of sight, and where the Lyman-$\alpha$ cross section is approximated as perfectly Lorentzian, Eq.~(\ref{eq:tau_DW_approx}) holds, or, in other words, $N_\mathrm{HI}^\mathrm{DW}$ encodes the same information as the pre-quasar damping wing optical depth $\tau_\mathrm{DW}^\mathrm{pre}(v = v_\mathrm{T})$ evaluated at the reference velocity offset $v_\mathrm{T}$.

We start by writing the integrand in Eq.~(\ref{eq:tau_DW_pre}) explicitly in terms of the integration variable $R$. This is required since we need to evaluate the Lyman-$\alpha$ cross section $\sigma_\alpha$ at wavelength
\begin{equation}
\label{eq:lambda_R}
    \lambda = \frac{1+z_\mathrm{QSO}}{1+z(R)}\,\lambda_\mathrm{rest}.
\end{equation}
The relation $z(R)$ between redshift $z$ and proper distance $R$ from the quasar can be obtained by inverting the light-travel distance relation
\begin{equation}
\label{eq:R_z}
    R(z) = \int_z^{z_\mathrm{QSO}} \frac{c}{(1+z')H(z')} \; \mathrm{d}z'.
\end{equation}
Certainly, the endpoint of reionization $z_\mathrm{end}$ puts an end to all contributions to the damping wing optical depth as per Eq.~(\ref{eq:tau_DW_pre}).\footnote{In fact, Section~\ref{sec:int_limits} led us to the even stronger conclusion that the damping wing imprint is only sensitive to the structure within the first $\sim 100\,\mathrm{cMpc}$ from the source.}
Since the integrand $c/(1+z')H(z')$ does not change significantly from $z_\mathrm{QSO}$ to $z_\mathrm{end}$, we can expand
Eq.~(\ref{eq:R_z}) to linear order around $z = z_\mathrm{QSO}$, finding
\begin{equation}
    R(z) \simeq \frac{c}{(1+z_\mathrm{QSO})\,H(z_\mathrm{QSO})} (z_\mathrm{QSO}-z).
\end{equation}
This relation can now easily be inverted, and we obtain
\begin{equation}
\label{eq:z_R}
    1+z(R) = (1+z_\mathrm{QSO}) \left( 1 - \frac{H(z_\mathrm{QSO})}{c}\,R\right)
\end{equation}
If we now evaluate the Lyman-$\alpha$ rest-frame wavelength $\lambda_\mathrm{rest} = \lambda_\alpha\,(1+\frac{v}{c})$ at the velocity offset $v=v_\mathrm{T}$, recalling that $R_\mathrm{T} \equiv +v_\mathrm{T}/H(z_\mathrm{QSO})$, and using Eq.~(\ref{eq:z_R}), we immediately find that Eq.~(\ref{eq:lambda_R}) turns into
\begin{equation}
\label{eq:lambda_R_final}
    \lambda = \lambda_\alpha\,\frac{c+H(z_\mathrm{QSO})\,R_\mathrm{T}}{c-H(z_\mathrm{QSO})\,R}.
\end{equation}
With an expression at hand for $\lambda(R)$, we now proceed by substituting this relation into the following expression for the scattering cross section $\sigma_\alpha$ of the Lyman-$\alpha$ line:
\begin{equation}
\label{eq:cross_section}
    \sigma_\alpha(\lambda) = \frac{\pi e^2}{m_e c}\; f_\alpha \; \phi_\alpha(\lambda).
\end{equation}
Here, $f_\alpha\simeq 0.416$ is the Lyman-$\alpha$ oscillator strength, and $\phi_\alpha(\lambda)$ is the line profile function, commonly assumed as a Voigt profile. Away from the Lyman-$\alpha$ resonance itself, 
we can assume to good approximation that $\phi_\alpha(\lambda)$ is of Lorentzian shape \citep{draine2011, bach2015}, i.e.,
\begin{equation}
\label{eq:line_profile}
    \phi_\alpha(\lambda) = \frac{\gamma_\alpha\,\lambda_\alpha/(\pi c)}{(\lambda_\alpha/\lambda-1)^2+\gamma_\alpha^2},
\end{equation}
where $\gamma_\alpha \equiv \Gamma_\alpha\,\lambda_\alpha/(4\pi c)$ with the decay constant $\Gamma_\alpha=6.265\,\times\,10^8\,\mathrm{s}^{-1}$ of the Lyman-$\alpha$ transition. Evaluating $\lambda_\mathrm{rest}$ at the velocity offset $v_\mathrm{T}$ assures that $\lambda$ (as given by Eq.~(\ref{eq:lambda_R})) is sufficiently far away from the Lyman-$\alpha$ resonance $\lambda_\alpha$ such that $\phi_\alpha(\lambda)$ is not only to good approximation of Lorentzian shape, but we can also safely neglect the second term in the denominator of Eq.~(\ref{eq:line_profile}). Utilizing Eq.~(\ref{eq:lambda_R_final}), we can explicitly rewrite Eq.~(\ref{eq:cross_section}) for the Lyman-$\alpha$ scattering cross section as a function of $R$:
\begin{equation}
    \sigma_\alpha(\lambda(R)) \simeq \frac{e^2}{m_e c^2}\,f_\alpha\,\gamma_\alpha\,\lambda_\alpha\,\frac{(c/H(z_\mathrm{QSO})-R_\mathrm{T})^2}{(R+R_\mathrm{T})^2}.
\end{equation}
Substituting this back into Eq.~(\ref{eq:tau_DW_pre}) shows that our weighting function (Eq.~(\ref{eq:weighting})) exactly mimics the scaling in the integrand of the optical depth integral. If we now let $R_\mathrm{min} \to 0$ and $R_\mathrm{max} \to R(z_\mathrm{QSO})$, we formally arrive at Eq.~(\ref{eq:tau_DW_approx}), showing that in these limits, $N_\mathrm{HI}^\mathrm{DW}$ is an optimal summary of the pre-quasar damping wing optical depth $\tau_\mathrm{DW}^\mathrm{pre}$. By instead choosing the integration limits in the way discussed in Section~\ref{sec:int_limits}, we can ensure that an analogous version of Eq.~(\ref{eq:tau_DW_approx}) even holds for the \textit{post}-quasar optical depth $\tau_\mathrm{DW}$.

\bsp	%
\label{lastpage}
\end{document}